\documentclass[aps,prl,twocolumn,superscriptaddress,letterpaper]{revtex4-1}

\usepackage{amssymb}
\usepackage{amsmath}
\usepackage{bm}
\usepackage{graphicx}
\usepackage{color}

\begin{document}

\title{Continuum of Bound States in a Non-Hermitian Model}

\author{Qiang Wang}
\affiliation{Division of Physics and Applied Physics, School of Physical and Mathematical Sciences, Nanyang Technological University,
Singapore 637371, Singapore}

\author{Changyan Zhu}
\affiliation{Division of Physics and Applied Physics, School of Physical and Mathematical Sciences, Nanyang Technological University,
Singapore 637371, Singapore}

\author{Xu Zheng}
\affiliation{Division of Physics and Applied Physics, School of Physical and Mathematical Sciences, Nanyang Technological University,
Singapore 637371, Singapore}

\author{Haoran Xue}
\affiliation{Division of Physics and Applied Physics, School of Physical and Mathematical Sciences, Nanyang Technological University,
Singapore 637371, Singapore}

\author{Baile Zhang}
\email{blzhang@ntu.edu.sg}
\affiliation{Division of Physics and Applied Physics, School of Physical and Mathematical Sciences, Nanyang Technological University, Singapore 637371, Singapore}
\affiliation{Centre for Disruptive Photonic Technologies, Nanyang Technological University, Singapore, 637371, Singapore}

\author{Y. D. Chong}
\email{yidong@ntu.edu.sg}
\affiliation{Division of Physics and Applied Physics, School of Physical and Mathematical Sciences, Nanyang Technological University, Singapore 637371, Singapore}
\affiliation{Centre for Disruptive Photonic Technologies, Nanyang Technological University, Singapore, 637371, Singapore}

\begin{abstract}
  In a Hermitian system, bound states must have quantized energies, whereas free states can form a continuum.  We demonstrate how this principle fails for non-Hermitian systems, by analyzing non-Hermitian continuous Hamiltonians with an imaginary momentum and Landau-type vector potential.  The eigenstates, which we call ``continuum Landau modes'' (CLMs), have gaussian spatial envelopes and form a continuum filling the complex energy plane.  We present experimentally-realizable 1D and 2D lattice models that host CLMs; the lattice eigenstates are localized and have other features matching the continuous model.  One of these lattices can serve as a rainbow trap, whereby the response to an excitation is concentrated at a position proportional to the frequency.  Another lattice can act a wave funnel, concentrating an input excitation onto a boundary over a wide frequency bandwidth.  Unlike recent funneling schemes based on the non-Hermitian skin effect, this requires a simple lattice design with reciprocal couplings.
\end{abstract}

\maketitle

The bound states of a quantum particle in an infinite continuous space have energies that are quantized \cite{ruelle1969remark, AmreinJan1974}.  This stems from a theorem that compact Hermitian Hamiltonians have pure point spectra \cite{rudin1991functional, teschl2009mathematical, conway2019course}, and accounts for the energy quantization in standard models such as the harmonic oscillator and potential well \cite{landau2013quantum}.  The principle also applies to Anderson localized states in random potentials, whose eigenenergies are dense but countable \cite{Frohlich1984, hundertmark2008short}, and to the long-wavelength regime of discrete (e.g., tight-binding) lattices \cite{luscher1982topology, symanzik1983continuum, michael1987towards, wen1990topological, mccann2006landau, zhang2006landau}.  Free states, by contrast, are spatially extended and form continuous spectra.  It is interesting to ask whether any physical system could behave differently, such as having an uncountable set of bound states.  How might such a Hamiltonian be realized, and what interesting properties might its eigenstates have?

Over the past two decades, a large literature has developed around the study of non-Hermitian Hamiltonians \cite{bender1998real, mostafazadeh2002P1, mostafazadeh2002P2, mostafazadeh2002P3, mostafazadeh2004physical, gong2018topological, kawabata2019symmetry, bergholtz2021exceptional}, catalyzed by the realization that such Hamiltonians can be implemented on synthetic classical wave structures such as photonic resonators and waveguide arrays \cite{guo2009observation, ruter2010observation, hodaei2014parity, feng2014single, schumer2022topological}.  Non-Hermitian systems have been found to exhibit various interesting and useful features with no Hermitian counterparts.  For instance, their spectra can contain exceptional points corresponding to the coalescence of multiple eigenstates \cite{heiss2012physics, zhen2015spawning, Xu2017, cerjan2019, miri2019exceptional}, which can be used to enhance optical sensing \cite{hodaei2017, chen2017exceptional, hokmabadi2019non}.  Another example is the non-Hermitian skin effect, whereby a non-Hermitian lattice's eigenstates condense onto its boundaries \cite{hatano1996localization, hatano1997vortex, lee2016anomalous, kunst2018biorthogonal, yao2018edge, yokomizo2019non, borgnia2020non, okuma2020topological, zhang2020correspondence, zhang2022universal}, with possible applications including light funneling \cite{weidemann2020topological} and the stabilization of laser modes \cite{Longhi2018, zhu2022anomalous}.  

This raises the possibility of using non-Hermitian systems to violate the standard distinctions between quantized bound states and continuous free states, which were derived under the assumption of Hermiticity \cite{SM}.  Here, we investigate a non-Hermitian Hamiltonian that has spatially localized energy eigenstates, which we call ``continuum Landau modes'' (CLMs), at every complex energy $E$. The Hamiltonian features a first-order \textit{imaginary} dependence on momentum, along with a Landau-type vector potential \cite{landau2013quantum}; its eigenstates, the CLMs, map to the zero modes of a continuous family of Hermitian 2D Dirac models \cite{Peres2006, ZhangGraphene2006, Kentaro2006, guinea2010energy, rechtsman2013strain, schomerus2013parity}.  In two dimensions (2D), the CLM's center position $\mathbf{r}_0$ varies linearly and continuously with the complex plane coordinates of $E$. By contrast, previous studies of non-Hermitian models with vector potentials found only quantized bound states, similar to the Hermitian case \cite{shen2018quantum, lu2021magnetic, shao2022cyclotron}.  We moreover show that the desired Hamiltonian arises in the long-wavelength limit of a 2D lattice with nonuniform complex mass and nonreciprocal hoppings \cite{yao2018edge, okuma2020topological, zhang2020correspondence, zhang2022universal}, which can be realized experimentally with photonic structures \cite{zhu2020photonic,song2020two} or other classical wave metamaterials \cite{helbig2020generalized, weidemann2020topological, zou2021observation, zhang2021observation, zhang2021acoustic, gao2022non, liang2022dynamic, Wang2022}.  If the lattice size is finite, the CLMs become countable but retain other key properties like the dependence between $\mathbf{r}_0$ and $E$.

One dimensional (1D) versions of the model can be realized in lattices with non-uniform real mass and nonreciprocal hoppings, with the CLM positions proportional to $\mathrm{Re}(E)$; or non-uniform imaginary mass and reciprocal hoppings, with CLM positions proportional to $\mathrm{Im}(E)$.  The first type of lattice can act as a non-Hermitian rainbow trap \cite{tsakmakidis2007trapped, gan2009rainbow, lu2021topological, lu2022chip}, in which excitations induce intensity peaks at positions proportional to the frequency.  Compared to a recent proposal for rainbow trapping using topological states within a bandgap \cite{lu2021topological, lu2022chip}, the CLM-based rainbow trapping scheme has the potential to operate over a wide frequency bandwidth.  The second type of 1D lattice acts as a wave funnel \cite{weidemann2020topological}: the response to an excitation is concentrated at one boundary.  This is similar to the funneling caused by the non-Hermitian skin effect \cite{weidemann2020topological}, but requires only the placement of onsite gain/loss without nonreciprocal couplings, and may therefore be easier to implement \cite{peng2014parity, chang2014parity, zhao2018topological, zhao2019non}.

\begin{figure}
\centering
\includegraphics[width=0.45\textwidth]{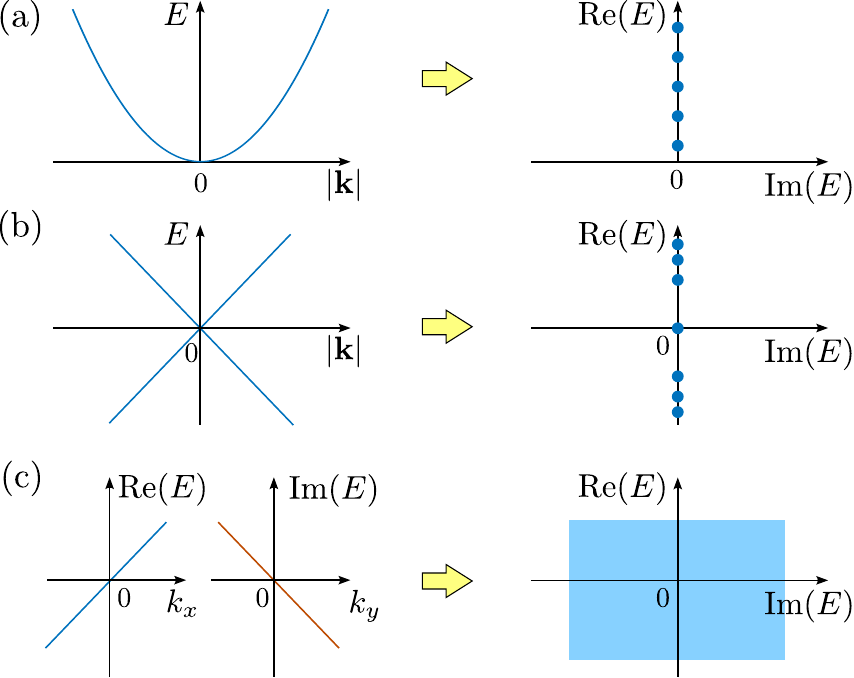}
\caption{Effects of a uniform magnetic field on the spectra of 2D models.  (a) For a nonrelativistic particle with quadratic dispersion, the spectrum collapses into discrete Landau levels.  (b) For a Dirac particle, the spectrum forms an unbounded sequence of Landau levels.  (c) For the non-Hermitian Hamiltonian \eqref{HamiltonainS}, the complex linear dispersion relation turns into a continuum of bound states filling the complex energy plane.}
\label{fig:f0}
\end{figure}

We begin by reviewing Landau quantization in 2D Hermitian systems.  As shown in Fig.~\ref{fig:f0}(a), a free nonrelativistic particle has a continuum of extended (free) states with a quadratic energy dispersion, and when a uniform magnetic field is applied, the spectrum collapses into a discrete set of Landau levels \cite{landau2013quantum}.  For a Dirac particle with a linear dispersion relation, a magnetic field likewise produces a discrete spectrum \cite{berestetskii1982quantum, Peres2006, ZhangGraphene2006, Kentaro2006}, as shown in Fig.~\ref{fig:f0}(b).  With appropriate gauge choices, the Landau levels are spanned by normalizable eigenfunctions (bound states) interpretable as cyclotron orbits.

Now consider the non-Hermitian 2D Hamiltonian
\begin{equation}
  \mathcal{H} =
  s_x \left[-i \frac{\partial}{\partial x} - By\right]
  + i s_y \left[-i \frac{\partial}{\partial y} + Bx \right],
  \label{HamiltonainS}
\end{equation}
where $s_{x,y}=\pm 1$ (these are scalars, not matrices).  For $B = 0$, this kind of ``non-Hermitian Dirac Hamiltonian'' has recently been analyzed in studies of non-Hermitian band topology \cite{kawabata2021topological, denner2021exceptional}; its spectrum is given by $E_0(\mathbf{k}) = s_x k_x + is_y k_y$, as shown in Fig.~\ref{fig:f0}(c).  The $B\ne 0$ case introduces a symmetric-gauge vector potential corresponding to a uniform out-of-plane magnetic field $2B \hat{z}$, via the substitution $-i\nabla \rightarrow -i\nabla + \mathbf{A}$.  The eigenstates of $\mathcal{H}$ are
\begin{align}
  \psi(x,y) &= C \exp\big[-\tau\, |\mathbf{r} - \mathbf{r}_0|^2
  + i \mathbf{q} \cdot \mathbf{r} \big], \label{Solution} \\
  \mathbf{r}_0(E, \mathbf{q})
  &= \frac{1}{B}\begin{pmatrix}
    ~\,\,\mathrm{Im}[E-E_0(\mathbf{q})]/s_y \\
    - \mathrm{Re}[E-E_0(\mathbf{q})]/s_x
    \end{pmatrix},
  \label{r0}
\end{align}
where $C$ is a normalization constant, $\tau = - s_xs_y B / 2$, $\mathbf{q} = (q_x, q_y)$ is an arbitrary real vector, and $E$ is the eigenenergy.  If $\tau > 0$, the wavefunctions are normalizable on $\mathbb{R}^2$ regardless of $E$ and $\mathbf{q}$, with characteristic length $\ell \sim B^{-1/2}$.  The eigenenergies fill the complex plane, as shown in Fig.~\ref{fig:f0}(c).  For each $E$, there is a continuum of bound states centered at different $\mathbf{r}_0$, via Eq.~\eqref{r0}; also, states with the same $\mathbf{q}$ but different $\mathbf{r}_0$ are non-orthogonal.  Note that such a continuum is not a generic consequence of non-Hermiticity; other recently-studied non-Hermitian models incorporating uniform magnetic fields exhibit the usual quantized spectra \cite{lu2021magnetic, shao2022cyclotron}.

We call these eigenstates CLMs because they are closely related to zeroth Landau level (0LL) modes of massless 2D Dirac fermions \cite{Peres2006, ZhangGraphene2006, Kentaro2006, guinea2010energy, rechtsman2013strain, schomerus2013parity}.  CLMs with a given energy $E$ have a one-to-one map with the 0LL modes of a given Hermitian Dirac Hamiltonian, whose gauge is determined by $E$.  The full set of CLM eigenstates for $\mathcal{H}$ thus maps to the 0LL modes of a \textit{family} of Dirac Hamiltonians with different gauges, and the CLMs are uncountable because the gauge can be continuously varied. For details about this mapping, including the role of gauge invariance, see the Supplemental Materials \cite{SM}.

Due to the localization of the CLMs, wavefunctions resist diffraction when undergoing time evolution with $\mathcal{H}$.  For instance, a gaussian wavepacket maintains its width under time evolution (even if the width differs from that of the CLMs); however, depending on the initial settings, the wavepacket can move and undergo amplification or decay, as detailed in the Supplemental Materials \cite{SM}.

\begin{figure*}
\centering
\includegraphics[width=0.90\textwidth]{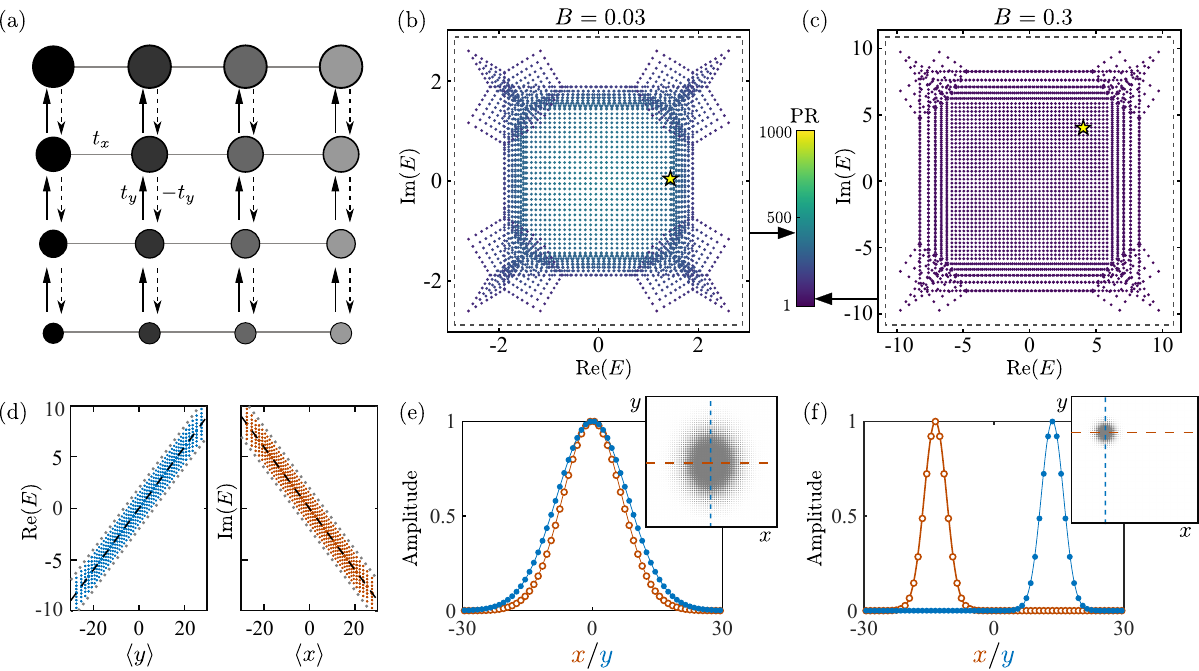}
\caption{Continuum Landau modes (CLMs) in a 2D lattice. (a) Schematic of a square lattice with reciprocal hoppings along $x$ (gray lines), nonreciprocal hoppings along $y$ (black arrows), and onsite mass $m_{x,y} = B(y-ix)$ (size and darkness of the circles indicate the real and imaginary parts).  (b)--(c) Complex energy spectra for finite lattices with (b) $B=0.03$ and (c) $B = 0.3$.  The color of each dot corresponds to the participation ratio (PR) of the eigenstate; a more localized state has lower PR.  The arrows on the color bar indicate the highest PR for the CLM ansatz, for each $B$.  The dashed boxes are the bounds on CLM eigenenergies derived from Eq.~\eqref{PosEn}.  (d) Plot of $\textrm{Re}(E)$ versus $\langle y\rangle$ (left panel) and $\textrm{Im}(E)$ versus $\langle x\rangle$ (right panel) for $B=0.3$.  The black dashes and gray dotted lines respectively indicate the theoretical central trend line (corresponding to $E^0_{\mathbf{k}+\mathbf{q}} \rightarrow 0$) and bounding lines derived from Eq.~\eqref{PosEn}.  (e)--(f) Wavefunction amplitude $|\psi_{\mathbf{r}}|$ for the eigenstates marked by yellow stars in (b) and (c) respectively.  Hollow and filled circles respectively indicate the variation with $x$ and $y$, along lines passing through the center of each gaussian; solid curves show the CLM predictions.  Insets show the distribution in the 2D plane.  In (b)--(f), we use $t_x=t_y=1$ and a lattice size of $60 \times 60$, with open boundary conditions.}
\label{fig:f1}
\end{figure*}

CLMs can also be observed in the continuum limit of discrete lattices.  Take the 2D lattice depicted in Fig.~\ref{fig:f1}(a), whose
Hamiltonian is
\begin{align}
  \begin{aligned}
  \mathcal{H} = \sum_{\mathbf{r}} & \Big[ B(y-ix)\, a_{\mathbf{r}}^\dagger a_{\mathbf{r}}
    + t_x \left(a_{\mathbf{r}-\hat{x}}^\dagger a_{\mathbf{r}} + \mathrm{h.c.}\right)  \\
    &\;+ t_y \left( a_{\mathbf{r}-\hat{y}}^\dagger a_{\mathbf{r}} - \mathrm{h.c.}\right)\Big],
  \end{aligned}
  \label{Lattice2D}
\end{align}
where $a^\dagger_{\mathbf{r}}$, $a_{\mathbf{r}}$ are creation and annihilation operators at $\mathbf{r} = (x,y)\in\mathbb{Z}^2$ (the lattice constant is set to 1), and $t_x, t_y \in \mathbb{R}$ are hopping coefficients.  $\mathcal{H}$ is non-Hermitian due to the imaginary part of the mass and the nonreciprocity of the $y$ hoppings \cite{yao2018edge, okuma2020topological, zhang2020correspondence, zhang2022universal}.  Such nonreciprocal hoppings can be realized on experimental platforms such as circuit lattices, fiber loops, and ring resonator lattices \cite{helbig2020generalized, weidemann2020topological, zou2021observation, zhang2021observation, zhang2021acoustic, gao2022non, liang2022dynamic, Wang2022}, which have notably been used to study the non-Hermitian skin effect \cite{hatano1996localization, hatano1997vortex, yao2018edge, yokomizo2019non, okuma2020topological, zhang2020correspondence, zhang2022universal}.  Note, however, that this lattice does not exhibit the skin effect \cite{SM}.

For $B = 0$, $\mathcal{H}$ has discrete translational symmetry and dispersion relation $E_{\mathbf{k}}^0 = 2(t_x\cos k_x+it_y\sin k_y)$, where $k_{x,y} \in [-\pi, \pi]$.  Taking $|\psi_{\mathbf{k}}\rangle = \sum_{\mathbf{r}}\exp(i\mathbf{k}\cdot\mathbf{r}) \Psi_{\mathbf{r}} a^\dagger_{\mathbf{r}}|\varnothing\rangle$, where $|\varnothing\rangle$ is the vacumm state, the slowly-varying envelope obeys $H_{\mathbf{k}} \Psi_{\mathbf{r}} = E \Psi_{\mathbf{r}}$, where \cite{SM}
\begin{align}
  H_{\mathbf{k}} &= E_{\mathbf{k}}^0
  - \left[-i \mu_{\mathbf{k}} \frac{\partial}{\partial x} - By \right]
  + i\left[-i\nu_{\mathbf{k}} \frac{\partial}{\partial y} - Bx\right],
  \label{Lattice2D1} \\
  \mu_{\mathbf{k}} &= 2t_x\sin k_x, \quad \nu_{\mathbf{k}} = 2t_y\cos k_y,
  \label{Lattice2D2} 
\end{align}
to first order in spatial derivatives.  Note that the complex masses in \eqref{Lattice2D} produce the pseudo vector potential \cite{Peres2006, ZhangGraphene2006, Kentaro2006, guinea2010energy, rechtsman2013strain, schomerus2013parity}.  For $B/\mu_{\mathbf{k}} < 0, B/\nu_{\mathbf{k}} > 0$, there exist CLMs similar to \eqref{Solution}, with $\tau$ replaced by $\tau_x = -B / 2\mu_{\mathbf{k}}$, $\tau_y = B / 2\nu_{\mathbf{k}}$, and Eq.~\eqref{r0} replaced by
\begin{align}
  \mathbf{r}_0(E,\mathbf{k},\mathbf{q}) =
  \frac{1}{B} \begin{pmatrix}
    -\mathrm{Im}\big(E-E^0_{\mathbf{k}+\mathbf{q}}\big) \\
    \;\;\;\mathrm{Re}\big(E-E^0_{\mathbf{k}+\mathbf{q}}\big)
  \end{pmatrix} +O(|\mathbf{q}|^2).
  \label{PosEn}
\end{align}
As $\Psi_{\mathrm{r}}$ is assumed to vary slowly in $\mathbf{r}$, the solutions are limited to the regime $|\mathbf{q}| \ll 1$.  If the lattice is infinite, they form a continuous set spanning all $E \in \mathbb{C}$.  For a finite lattice, the eigenstates are finite and hence countable, and the CLMs reduce to a band over a finite area in the $E$ plane.  In Fig.~\ref{fig:f1}(b)--(c), we plot the spectra for $B = 0.03, 0.3$, each lattice having size $L_x = L_y = 60$, open boundary conditions, and $t_x = t_y = 1$.  By requiring $\mathbf{r}_0$ to lie in the lattice, Eq.~\eqref{PosEn} implies the bounds $|\mathrm{Re}(E)| \lesssim BL_y/2+2t_x$ and $|\mathrm{Im}(E)| \lesssim BL_x/2+2t_y$.  For large $L_{x,y}$, the energy discretization is of order $B$.

All of the numerically obtained eigenstates are CLMs.  The color of each data point in Fig.~\ref{fig:f1}(b)--(c) indicates the participation ratio (PR), defined for a wavefunction $\psi_{\mathbf{r}} = \langle\mathbf{r}|\psi\rangle$ as $\langle \psi|\psi\rangle^2 / \sum_{\mathbf{r}} |\langle \mathbf{r}|\psi\rangle|^4$, with large PR (of order $N^2$) corresponding to extended states \cite{thouless1974}.  We find that all eigenstates have PR consistent with the CLM predictions, and well below $N^2$ (for each case, the maximum PR, attained when $|\mu_{\mathbf{k}}|=|\nu_{\mathbf{k}}|=1$, is indicated by an arrow in the color bar). Fig.~\ref{fig:f1}(d) plots each eigenstate's energy against the position expectation values $\langle x\rangle$ and $\langle y\rangle$, for $B=0.3$.  This reveals the linear relationship between $E$ and $\mathbf{r}_0$ predicted in Eq.~\eqref{PosEn} (indicated by dashes), and the upper and lower bounds introduced by the $E^0_{\mathbf{k}+\mathbf{q}}$ term (dotted lines).  Fig.~\ref{fig:f1}(e)--(f) compares the spatial amplitude $|\psi_{\mathbf{r}}|$ for two arbitrarily-chosen numerical eigenstates to the CLM solutions, which are in good agreement.

\begin{figure}
\centering
\includegraphics[width=0.483\textwidth]{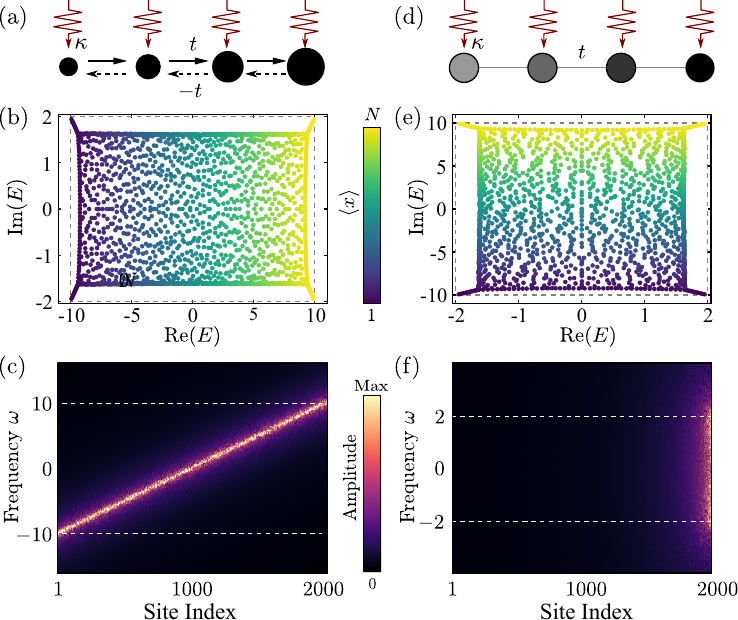}
\caption{Rainbow trapping and wave funneling in 1D lattices. (a) 1D lattice with nonreciprocal nearest neighbor hoppings of $t$ (solid arrows) and $-t$ (dashed arrows).  Each site $j$ has real mass $m_j = B(j-j_0)$, where $j \in[1, N]$ and $j_0=(N+1)/2$. (b) Complex energy spectrum for the lattice in (a) with $N = 2000$, $t = 1$, and $B = 0.01$. The color of each dot indicates the eigenstate's position expectation value $\langle x\rangle$.  The dashed box shows the CLM energy bounds described in the text.  (c) Site-dependent amplitudes under steady state excitation at frequency $\omega$ for the lattice in (b), with an additional per-site damping term $\Delta m=1.9i$ to avoid blowup.  On each site, the excitation has uniform amplitude but a random phase drawn uniformly from $[0,2\pi)$.  The two dashes show the working band $[-BN/2,BN/2]$.  The peak position found to be proportional to $\omega$. (d) 1D lattice with reciprocal nearest neighbor hopping $t$, and onsite mass $m_j=iB(j-j_0)$. (e) Complex energy spectrum for the lattice in (d) with $N = 2000$, $t = 1$, and $B = 0.01$.  (f) Site-dependent amplitudes under the same excitation as in (c), using the lattice in (e) with additional per-site damping $\Delta m = 9.9i$.  Dashes show the working band $[-2t,2t]$.  Funneling toward the large-$j$ boundary is observed.  In (c) and (f), the coupling of the excitation to each site is $\kappa=0.2$.}
\label{fig:f2}
\end{figure}

CLMs can also be realized in 1D lattices, which are simpler to implement experimentally.  We study two kinds of lattices, which have different behaviors.

The first 1D lattice, shown in Fig.~\ref{fig:f2}(a), has a real mass gradient and nonreciprocal hoppings.  Its Hamiltonian is
\begin{align}
\mathcal{H}
&= \sum_{j} \left[ B(j- j_0)\, a_j^\dagger a_j
+t a_j^\dagger a_{j-1}-t a_{j-1}^\dagger a_{j} \right],
\label{Lattice}
\end{align}
where $B, t\in \mathbb{R}$.  For a finite lattice with $1 \le j \le N$, we take $j_0 = \frac{1}{2}(N+1)$.  The nonreciprocal nearest neighbor hoppings $\pm t \in \mathbb{R}$ are indicated by solid and dashed arrows.  For $B=0$, the spectrum $E_k^0 =2it\sin k$ (where $k\in[-\pi,\pi]$) is purely imaginary, lacking a point gap, and the lattice does not exhibit the non-Hermitian skin effect \cite{okuma2020topological, zhang2020correspondence}. For $B\neq 0$, the effective Hamiltonian can be derived via the same procedures as in Eqs.~\eqref{Lattice2D}--\eqref{Lattice2D2}:
\begin{equation}
  H_{k} = E_{k}^0 + Bx + 2t \cos k\; \frac{\partial{}}{\partial{x}}.
\label{Hamiltonain3}
\end{equation}
This has CLM solutions\cite{SM} for $B/(t\cos {k})<0$ (i.e., $k$ in half of the Brillouin zone).  As in the 2D model, $E$ takes any value in $\mathbb{C}$, but for finite $N$ the eigenvalues reduce to a band as shown in Fig.~\ref{fig:f2}(b), bounded by $|\mathrm{Re}(E)| \lesssim BN/2$ and $|\mathrm{Im}(E)| \lesssim 2t$.  The bandwidth $\Delta[\mathrm{Re}(E)] \sim BN$ is the detuning between the two end sites.  The CLM center positions are $x_0 = \mathrm{Re}(E)/B$ (independent of $k$, since $E_{k}^0$ is imaginary), consistent with the colors in Fig.~\ref{fig:f2}(b).

Systems that have eigenstates localized at positions proportional to frequency can serve as ``rainbow traps'' \cite{tsakmakidis2007trapped, gan2009rainbow, lu2021topological, lu2022chip}, with potential applications in wave buffering and frequency demultiplexing. To demonstrate this, we apply a spatially incoherent monochromatic excitation $F(t)= e^{-iwt} \cdot [e^{i\theta_1}, \dots, e^{i\theta_N}]^T$, where each $\theta_j$ is a random initial phase.  We then use the Green's function to calculate the steady-state response \cite{kadanoff2018quantum, SM}. As shown in Fig.~\ref{fig:f2}(c), the response has a sharp amplitude peak positioned proportional to the excitation frequency.  Previously-studied rainbow traps have been based on Hermitian systems, and operate on different principles.  The present scheme relies on the properties of the CLMs; when a comparable non-Hermitian lattice lacking CLMs is excited, the intensity peaks are distributed in different positions along the lattice with no evident relationship with frequency (see Supplemental Materials \cite{SM}).  The bandwidth $\Delta E \sim BN$ can be increased via the mass gradient $B$ or lattice size $N$.

Another 1D lattice supporting CLMs, depicted in Fig.~\ref{fig:f2}(d), has reciprocal hoppings along with a gradient in the onsite gain/loss.  The Hamiltonian is
\begin{equation}
\mathcal{H} = \sum_j \left[ iB(j-j_0) a_j^\dagger a_j+t \left( a_j^\dagger a_{j-1}+ a_{j-1}^\dagger a_{j} \right) \right].
\label{Lattice2}
\end{equation}
For $B=0$, this is Hermitian and the spectrum is $E_k^0 =2t\cos k$.  For $B\neq 0$, we obtain
\begin{equation}
  H_{k} = E_{k}^0 + iBx + i 2t \sin k\; \frac{\partial{}}{\partial{x}}.
\label{Hamiltonain4}
\end{equation}
There are CLM solutions\cite{SM} for $B/(t\sin {k})<0$. Fig.~\ref{fig:f2}(e) shows the spectrum for a finite lattice, which is bounded by $|\mathrm{Re}(E)| \lesssim 2t$ and $|\mathrm{Im}(E)| \lesssim BN/2$.  As the CLMs are centered at $x_0 = \mathrm{Im}(E)/B$, the modes with highest relative gain occur at a boundary.  Under a spatially incoherent excitation, the steady-state response is concentrated at the boundary, as shown in Fig.~\ref{fig:f2}(f).  We emphasize that this is not simply due to the boundary site having the highest relative gain, since the funneling effect occurs over a bandwidth of $\Delta[\mathrm{Re}(E)] \sim 4t$, rather than the narrow linewidth of an isolated resonance.  Similar funneling behavior has been associated with the non-Hermitian skin effect \cite{hatano1996localization, hatano1997vortex, yao2018edge, yokomizo2019non, weidemann2020topological, okuma2020topological, zhang2020correspondence, zhang2022universal}, but the present lattice does not exhibit the skin effect\cite{SM}.  This way of implementing a wave funnel requires neither nonreciprocal hoppings nor complicated lattice symmetries.  In the Supplemental Materials, we show that a similar lattice lacking CLMs does not induce funneling \cite{SM}.

In conclusion, we have shown that a non-Hermitian Hamiltonian can host an uncountably infinite set of bound states, violating the intuition that bound states should be quantized.  The bound states possess gaussian envelopes and are related to zeroth Landau level modes \cite{Peres2006, ZhangGraphene2006, Kentaro2006, guinea2010energy, rechtsman2013strain, schomerus2013parity}.  We have shown how to implement these eigenstates in 1D and 2D lattices that are experimentally accessible via classical metamaterial platforms \cite{helbig2020generalized, weidemann2020topological, zou2021observation, zhang2021observation, zhang2021acoustic, gao2022non, liang2022dynamic, Wang2022}.  In 1D lattices, the linear relationship between position and energy allows for non-Hermitian rainbow trapping and light funneling functionalities. In future work, it would be interesting to determine the general conditions under which non-quantized bound states can arise, and to explore other ways to violate the distinction between bound and free states.  Continuous families of bound states might emerge among other non-Hermitian systems unrelated to zeroth Landau level modes, with properties different from those studied here.

This work was supported by the Singapore MOE Academic Research Fund Tier~3 Grant MOE2016-T3-1-006 and Tier~1 Grant RG148/20, and by the National Research Foundation Competitive Research Programs NRF-CRP23-2019-0005 and NRF-CRP23-2019-0007.

\bibliography{citepaper}

\begin{thebibliography}{80}%
\makeatletter
\providecommand \@ifxundefined [1]{%
 \@ifx{#1\undefined}
}%
\providecommand \@ifnum [1]{%
 \ifnum #1\expandafter \@firstoftwo
 \else \expandafter \@secondoftwo
 \fi
}%
\providecommand \@ifx [1]{%
 \ifx #1\expandafter \@firstoftwo
 \else \expandafter \@secondoftwo
 \fi
}%
\providecommand \natexlab [1]{#1}%
\providecommand \enquote  [1]{``#1''}%
\providecommand \bibnamefont  [1]{#1}%
\providecommand \bibfnamefont [1]{#1}%
\providecommand \citenamefont [1]{#1}%
\providecommand \href@noop [0]{\@secondoftwo}%
\providecommand \href [0]{\begingroup \@sanitize@url \@href}%
\providecommand \@href[1]{\@@startlink{#1}\@@href}%
\providecommand \@@href[1]{\endgroup#1\@@endlink}%
\providecommand \@sanitize@url [0]{\catcode `\\12\catcode `\$12\catcode
  `\&12\catcode `\#12\catcode `\^12\catcode `\_12\catcode `\%12\relax}%
\providecommand \@@startlink[1]{}%
\providecommand \@@endlink[0]{}%
\providecommand \url  [0]{\begingroup\@sanitize@url \@url }%
\providecommand \@url [1]{\endgroup\@href {#1}{\urlprefix }}%
\providecommand \urlprefix  [0]{URL }%
\providecommand \Eprint [0]{\href }%
\providecommand \doibase [0]{http://dx.doi.org/}%
\providecommand \selectlanguage [0]{\@gobble}%
\providecommand \bibinfo  [0]{\@secondoftwo}%
\providecommand \bibfield  [0]{\@secondoftwo}%
\providecommand \translation [1]{[#1]}%
\providecommand \BibitemOpen [0]{}%
\providecommand \bibitemStop [0]{}%
\providecommand \bibitemNoStop [0]{.\EOS\space}%
\providecommand \EOS [0]{\spacefactor3000\relax}%
\providecommand \BibitemShut  [1]{\csname bibitem#1\endcsname}%
\let\auto@bib@innerbib\@empty
\bibitem [{\citenamefont {Ruelle}(1969)}]{ruelle1969remark}%
  \BibitemOpen
  \bibfield  {author} {\bibinfo {author} {\bibfnamefont {D.}~\bibnamefont
  {Ruelle}},\ }\href@noop {} {\bibfield  {journal} {\bibinfo  {journal} {Il
  Nuovo Cimento A (1965-1970)}\ }\textbf {\bibinfo {volume} {61}},\ \bibinfo
  {pages} {655} (\bibinfo {year} {1969})}\BibitemShut {NoStop}%
\bibitem [{\citenamefont {Amrein}\ and\ \citenamefont
  {Georgescu}(1974)}]{AmreinJan1974}%
  \BibitemOpen
  \bibfield  {author} {\bibinfo {author} {\bibfnamefont {W.~O.}\ \bibnamefont
  {Amrein}}\ and\ \bibinfo {author} {\bibfnamefont {V.}~\bibnamefont
  {Georgescu}},\ }\href@noop {} {\bibfield  {journal} {\bibinfo  {journal}
  {Helv. Phys. Acta}\ }\textbf {\bibinfo {volume} {46}},\ \bibinfo {pages}
  {635} (\bibinfo {year} {1974})}\BibitemShut {NoStop}%
\bibitem [{\citenamefont {Rudin}(1991)}]{rudin1991functional}%
  \BibitemOpen
  \bibfield  {author} {\bibinfo {author} {\bibfnamefont {W.}~\bibnamefont
  {Rudin}},\ }\href {https://books.google.com.sg/books?id=Sh\_vAAAAMAAJ} {\emph
  {\bibinfo {title} {Functional Analysis}}},\ International series in pure and
  applied mathematics\ (\bibinfo  {publisher} {McGraw-Hill},\ \bibinfo {year}
  {1991})\BibitemShut {NoStop}%
\bibitem [{\citenamefont {Teschl}(2009)}]{teschl2009mathematical}%
  \BibitemOpen
  \bibfield  {author} {\bibinfo {author} {\bibfnamefont {G.}~\bibnamefont
  {Teschl}},\ }\href@noop {} {\bibfield  {journal} {\bibinfo  {journal}
  {Graduate Studies in Mathematics}\ }\textbf {\bibinfo {volume} {99}},\
  \bibinfo {pages} {106} (\bibinfo {year} {2009})}\BibitemShut {NoStop}%
\bibitem [{\citenamefont {Conway}(2019)}]{conway2019course}%
  \BibitemOpen
  \bibfield  {author} {\bibinfo {author} {\bibfnamefont {J.~B.}\ \bibnamefont
  {Conway}},\ }\href@noop {} {\emph {\bibinfo {title} {A course in functional
  analysis}}},\ Vol.~\bibinfo {volume} {96}\ (\bibinfo  {publisher}
  {Springer},\ \bibinfo {year} {2019})\BibitemShut {NoStop}%
\bibitem [{\citenamefont {Landau}\ and\ \citenamefont
  {Lifshitz}(2013)}]{landau2013quantum}%
  \BibitemOpen
  \bibfield  {author} {\bibinfo {author} {\bibfnamefont {L.~D.}\ \bibnamefont
  {Landau}}\ and\ \bibinfo {author} {\bibfnamefont {E.~M.}\ \bibnamefont
  {Lifshitz}},\ }\href@noop {} {\emph {\bibinfo {title} {Quantum mechanics:
  non-relativistic theory}}},\ Vol.~\bibinfo {volume} {3}\ (\bibinfo
  {publisher} {Elsevier},\ \bibinfo {year} {2013})\BibitemShut {NoStop}%
\bibitem [{\citenamefont {Fr{\"o}hlich}\ and\ \citenamefont
  {Spencer}(1984)}]{Frohlich1984}%
  \BibitemOpen
  \bibfield  {author} {\bibinfo {author} {\bibfnamefont {J.}~\bibnamefont
  {Fr{\"o}hlich}}\ and\ \bibinfo {author} {\bibfnamefont {T.}~\bibnamefont
  {Spencer}},\ }\href {\doibase https://doi.org/10.1016/0370-1573(84)90061-9}
  {\bibfield  {journal} {\bibinfo  {journal} {Phys. Rep.}\ }\textbf {\bibinfo
  {volume} {103}},\ \bibinfo {pages} {9} (\bibinfo {year} {1984})}\BibitemShut
  {NoStop}%
\bibitem [{\citenamefont {Hundertmark}(2008)}]{hundertmark2008short}%
  \BibitemOpen
  \bibfield  {author} {\bibinfo {author} {\bibfnamefont {D.}~\bibnamefont
  {Hundertmark}},\ }in\ \href@noop {} {\emph {\bibinfo {booktitle} {Analysis
  and stochastics of growth processes and interface models}}},\ \bibinfo
  {editor} {edited by\ \bibinfo {editor} {\bibfnamefont {P.}~\bibnamefont
  {M{\"o}rters}}, \bibinfo {editor} {\bibfnamefont {R.}~\bibnamefont {Moser}},
  \bibinfo {editor} {\bibfnamefont {M.}~\bibnamefont {Penrose}}, \bibinfo
  {editor} {\bibfnamefont {H.}~\bibnamefont {Schwetlick}}, \ and\ \bibinfo
  {editor} {\bibfnamefont {J.}~\bibnamefont {Zimmer}}}\ (\bibinfo  {publisher}
  {Oxford University Press},\ \bibinfo {address} {Oxford},\ \bibinfo {year}
  {2008})\ Chap.~\bibinfo {chapter} {9}, p.\ \bibinfo {pages} {194}\BibitemShut
  {NoStop}%
\bibitem [{\citenamefont {L{\"u}scher}(1982)}]{luscher1982topology}%
  \BibitemOpen
  \bibfield  {author} {\bibinfo {author} {\bibfnamefont {M.}~\bibnamefont
  {L{\"u}scher}},\ }\href@noop {} {\bibfield  {journal} {\bibinfo  {journal}
  {Comm. Math. Phys.}\ }\textbf {\bibinfo {volume} {85}},\ \bibinfo {pages}
  {39} (\bibinfo {year} {1982})}\BibitemShut {NoStop}%
\bibitem [{\citenamefont {Symanzik}(1983)}]{symanzik1983continuum}%
  \BibitemOpen
  \bibfield  {author} {\bibinfo {author} {\bibfnamefont {K.}~\bibnamefont
  {Symanzik}},\ }\href@noop {} {\bibfield  {journal} {\bibinfo  {journal}
  {Nucl. Phys. B}\ }\textbf {\bibinfo {volume} {226}},\ \bibinfo {pages} {187}
  (\bibinfo {year} {1983})}\BibitemShut {NoStop}%
\bibitem [{\citenamefont {Michael}\ and\ \citenamefont
  {Teper}(1987)}]{michael1987towards}%
  \BibitemOpen
  \bibfield  {author} {\bibinfo {author} {\bibfnamefont {C.}~\bibnamefont
  {Michael}}\ and\ \bibinfo {author} {\bibfnamefont {M.}~\bibnamefont
  {Teper}},\ }\href@noop {} {\bibfield  {journal} {\bibinfo  {journal} {Phys.
  Lett. B}\ }\textbf {\bibinfo {volume} {199}},\ \bibinfo {pages} {95}
  (\bibinfo {year} {1987})}\BibitemShut {NoStop}%
\bibitem [{\citenamefont {Wen}(1990)}]{wen1990topological}%
  \BibitemOpen
  \bibfield  {author} {\bibinfo {author} {\bibfnamefont {X.-G.}\ \bibnamefont
  {Wen}},\ }\href@noop {} {\bibfield  {journal} {\bibinfo  {journal} {Int. J.
  Mod. Phys. B}\ }\textbf {\bibinfo {volume} {4}},\ \bibinfo {pages} {239}
  (\bibinfo {year} {1990})}\BibitemShut {NoStop}%
\bibitem [{\citenamefont {McCann}\ and\ \citenamefont
  {Fal'ko}(2006)}]{mccann2006landau}%
  \BibitemOpen
  \bibfield  {author} {\bibinfo {author} {\bibfnamefont {E.}~\bibnamefont
  {McCann}}\ and\ \bibinfo {author} {\bibfnamefont {V.~I.}\ \bibnamefont
  {Fal'ko}},\ }\href@noop {} {\bibfield  {journal} {\bibinfo  {journal} {Phys.
  Rev. Lett.}\ }\textbf {\bibinfo {volume} {96}},\ \bibinfo {pages} {086805}
  (\bibinfo {year} {2006})}\BibitemShut {NoStop}%
\bibitem [{\citenamefont {Zhang}\ \emph
  {et~al.}(2006{\natexlab{a}})\citenamefont {Zhang}, \citenamefont {Jiang},
  \citenamefont {Small}, \citenamefont {Purewal}, \citenamefont {Tan},
  \citenamefont {Fazlollahi}, \citenamefont {Chudow}, \citenamefont {Jaszczak},
  \citenamefont {Stormer},\ and\ \citenamefont {Kim}}]{zhang2006landau}%
  \BibitemOpen
  \bibfield  {author} {\bibinfo {author} {\bibfnamefont {Y.}~\bibnamefont
  {Zhang}}, \bibinfo {author} {\bibfnamefont {Z.}~\bibnamefont {Jiang}},
  \bibinfo {author} {\bibfnamefont {J.}~\bibnamefont {Small}}, \bibinfo
  {author} {\bibfnamefont {M.}~\bibnamefont {Purewal}}, \bibinfo {author}
  {\bibfnamefont {Y.-W.}\ \bibnamefont {Tan}}, \bibinfo {author} {\bibfnamefont
  {M.}~\bibnamefont {Fazlollahi}}, \bibinfo {author} {\bibfnamefont
  {J.}~\bibnamefont {Chudow}}, \bibinfo {author} {\bibfnamefont
  {J.}~\bibnamefont {Jaszczak}}, \bibinfo {author} {\bibfnamefont
  {H.}~\bibnamefont {Stormer}}, \ and\ \bibinfo {author} {\bibfnamefont
  {P.}~\bibnamefont {Kim}},\ }\href@noop {} {\bibfield  {journal} {\bibinfo
  {journal} {Phys. Rev. Lett.}\ }\textbf {\bibinfo {volume} {96}},\ \bibinfo
  {pages} {136806} (\bibinfo {year} {2006}{\natexlab{a}})}\BibitemShut
  {NoStop}%
\bibitem [{\citenamefont {Bender}\ and\ \citenamefont
  {Boettcher}(1998)}]{bender1998real}%
  \BibitemOpen
  \bibfield  {author} {\bibinfo {author} {\bibfnamefont {C.~M.}\ \bibnamefont
  {Bender}}\ and\ \bibinfo {author} {\bibfnamefont {S.}~\bibnamefont
  {Boettcher}},\ }\href@noop {} {\bibfield  {journal} {\bibinfo  {journal}
  {Phys. Rev. Lett.}\ }\textbf {\bibinfo {volume} {80}},\ \bibinfo {pages}
  {5243} (\bibinfo {year} {1998})}\BibitemShut {NoStop}%
\bibitem [{\citenamefont
  {Mostafazadeh}(2002{\natexlab{a}})}]{mostafazadeh2002P1}%
  \BibitemOpen
  \bibfield  {author} {\bibinfo {author} {\bibfnamefont {A.}~\bibnamefont
  {Mostafazadeh}},\ }\href@noop {} {\bibfield  {journal} {\bibinfo  {journal}
  {J. Math. Phys.}\ }\textbf {\bibinfo {volume} {43}},\ \bibinfo {pages} {205}
  (\bibinfo {year} {2002}{\natexlab{a}})}\BibitemShut {NoStop}%
\bibitem [{\citenamefont
  {Mostafazadeh}(2002{\natexlab{b}})}]{mostafazadeh2002P2}%
  \BibitemOpen
  \bibfield  {author} {\bibinfo {author} {\bibfnamefont {A.}~\bibnamefont
  {Mostafazadeh}},\ }\href@noop {} {\bibfield  {journal} {\bibinfo  {journal}
  {J. Math. Phys.}\ }\textbf {\bibinfo {volume} {43}},\ \bibinfo {pages} {2814}
  (\bibinfo {year} {2002}{\natexlab{b}})}\BibitemShut {NoStop}%
\bibitem [{\citenamefont
  {Mostafazadeh}(2002{\natexlab{c}})}]{mostafazadeh2002P3}%
  \BibitemOpen
  \bibfield  {author} {\bibinfo {author} {\bibfnamefont {A.}~\bibnamefont
  {Mostafazadeh}},\ }\href@noop {} {\bibfield  {journal} {\bibinfo  {journal}
  {J. Math. Phys.}\ }\textbf {\bibinfo {volume} {43}},\ \bibinfo {pages} {3944}
  (\bibinfo {year} {2002}{\natexlab{c}})}\BibitemShut {NoStop}%
\bibitem [{\citenamefont {Mostafazadeh}\ and\ \citenamefont
  {Batal}(2004)}]{mostafazadeh2004physical}%
  \BibitemOpen
  \bibfield  {author} {\bibinfo {author} {\bibfnamefont {A.}~\bibnamefont
  {Mostafazadeh}}\ and\ \bibinfo {author} {\bibfnamefont {A.}~\bibnamefont
  {Batal}},\ }\href@noop {} {\bibfield  {journal} {\bibinfo  {journal} {J.
  Phys. A: Math. Gen.}\ }\textbf {\bibinfo {volume} {37}},\ \bibinfo {pages}
  {11645} (\bibinfo {year} {2004})}\BibitemShut {NoStop}%
\bibitem [{\citenamefont {Gong}\ \emph {et~al.}(2018)\citenamefont {Gong},
  \citenamefont {Ashida}, \citenamefont {Kawabata}, \citenamefont {Takasan},
  \citenamefont {Higashikawa},\ and\ \citenamefont
  {Ueda}}]{gong2018topological}%
  \BibitemOpen
  \bibfield  {author} {\bibinfo {author} {\bibfnamefont {Z.}~\bibnamefont
  {Gong}}, \bibinfo {author} {\bibfnamefont {Y.}~\bibnamefont {Ashida}},
  \bibinfo {author} {\bibfnamefont {K.}~\bibnamefont {Kawabata}}, \bibinfo
  {author} {\bibfnamefont {K.}~\bibnamefont {Takasan}}, \bibinfo {author}
  {\bibfnamefont {S.}~\bibnamefont {Higashikawa}}, \ and\ \bibinfo {author}
  {\bibfnamefont {M.}~\bibnamefont {Ueda}},\ }\href@noop {} {\bibfield
  {journal} {\bibinfo  {journal} {Phys. Rev. X}\ }\textbf {\bibinfo {volume}
  {8}},\ \bibinfo {pages} {031079} (\bibinfo {year} {2018})}\BibitemShut
  {NoStop}%
\bibitem [{\citenamefont {Kawabata}\ \emph {et~al.}(2019)\citenamefont
  {Kawabata}, \citenamefont {Shiozaki}, \citenamefont {Ueda},\ and\
  \citenamefont {Sato}}]{kawabata2019symmetry}%
  \BibitemOpen
  \bibfield  {author} {\bibinfo {author} {\bibfnamefont {K.}~\bibnamefont
  {Kawabata}}, \bibinfo {author} {\bibfnamefont {K.}~\bibnamefont {Shiozaki}},
  \bibinfo {author} {\bibfnamefont {M.}~\bibnamefont {Ueda}}, \ and\ \bibinfo
  {author} {\bibfnamefont {M.}~\bibnamefont {Sato}},\ }\href@noop {} {\bibfield
   {journal} {\bibinfo  {journal} {Phys. Rev. X}\ }\textbf {\bibinfo {volume}
  {9}},\ \bibinfo {pages} {041015} (\bibinfo {year} {2019})}\BibitemShut
  {NoStop}%
\bibitem [{\citenamefont {Bergholtz}\ \emph {et~al.}(2021)\citenamefont
  {Bergholtz}, \citenamefont {Budich},\ and\ \citenamefont
  {Kunst}}]{bergholtz2021exceptional}%
  \BibitemOpen
  \bibfield  {author} {\bibinfo {author} {\bibfnamefont {E.~J.}\ \bibnamefont
  {Bergholtz}}, \bibinfo {author} {\bibfnamefont {J.~C.}\ \bibnamefont
  {Budich}}, \ and\ \bibinfo {author} {\bibfnamefont {F.~K.}\ \bibnamefont
  {Kunst}},\ }\href@noop {} {\bibfield  {journal} {\bibinfo  {journal} {Rev.
  Mod. Phys.}\ }\textbf {\bibinfo {volume} {93}},\ \bibinfo {pages} {015005}
  (\bibinfo {year} {2021})}\BibitemShut {NoStop}%
\bibitem [{\citenamefont {Guo}\ \emph {et~al.}(2009)\citenamefont {Guo},
  \citenamefont {Salamo}, \citenamefont {Duchesne}, \citenamefont {Morandotti},
  \citenamefont {Volatier-Ravat}, \citenamefont {Aimez}, \citenamefont
  {Siviloglou},\ and\ \citenamefont {Christodoulides}}]{guo2009observation}%
  \BibitemOpen
  \bibfield  {author} {\bibinfo {author} {\bibfnamefont {A.}~\bibnamefont
  {Guo}}, \bibinfo {author} {\bibfnamefont {G.}~\bibnamefont {Salamo}},
  \bibinfo {author} {\bibfnamefont {D.}~\bibnamefont {Duchesne}}, \bibinfo
  {author} {\bibfnamefont {R.}~\bibnamefont {Morandotti}}, \bibinfo {author}
  {\bibfnamefont {M.}~\bibnamefont {Volatier-Ravat}}, \bibinfo {author}
  {\bibfnamefont {V.}~\bibnamefont {Aimez}}, \bibinfo {author} {\bibfnamefont
  {G.}~\bibnamefont {Siviloglou}}, \ and\ \bibinfo {author} {\bibfnamefont
  {D.}~\bibnamefont {Christodoulides}},\ }\href@noop {} {\bibfield  {journal}
  {\bibinfo  {journal} {Phys. Rev. Lett.}\ }\textbf {\bibinfo {volume} {103}},\
  \bibinfo {pages} {093902} (\bibinfo {year} {2009})}\BibitemShut {NoStop}%
\bibitem [{\citenamefont {R{\"u}ter}\ \emph {et~al.}(2010)\citenamefont
  {R{\"u}ter}, \citenamefont {Makris}, \citenamefont {El-Ganainy},
  \citenamefont {Christodoulides}, \citenamefont {Segev},\ and\ \citenamefont
  {Kip}}]{ruter2010observation}%
  \BibitemOpen
  \bibfield  {author} {\bibinfo {author} {\bibfnamefont {C.~E.}\ \bibnamefont
  {R{\"u}ter}}, \bibinfo {author} {\bibfnamefont {K.~G.}\ \bibnamefont
  {Makris}}, \bibinfo {author} {\bibfnamefont {R.}~\bibnamefont {El-Ganainy}},
  \bibinfo {author} {\bibfnamefont {D.~N.}\ \bibnamefont {Christodoulides}},
  \bibinfo {author} {\bibfnamefont {M.}~\bibnamefont {Segev}}, \ and\ \bibinfo
  {author} {\bibfnamefont {D.}~\bibnamefont {Kip}},\ }\href@noop {} {\bibfield
  {journal} {\bibinfo  {journal} {Nat. Phys.}\ }\textbf {\bibinfo {volume}
  {6}},\ \bibinfo {pages} {192} (\bibinfo {year} {2010})}\BibitemShut {NoStop}%
\bibitem [{\citenamefont {Hodaei}\ \emph {et~al.}(2014)\citenamefont {Hodaei},
  \citenamefont {Miri}, \citenamefont {Heinrich}, \citenamefont
  {Christodoulides},\ and\ \citenamefont {Khajavikhan}}]{hodaei2014parity}%
  \BibitemOpen
  \bibfield  {author} {\bibinfo {author} {\bibfnamefont {H.}~\bibnamefont
  {Hodaei}}, \bibinfo {author} {\bibfnamefont {M.-A.}\ \bibnamefont {Miri}},
  \bibinfo {author} {\bibfnamefont {M.}~\bibnamefont {Heinrich}}, \bibinfo
  {author} {\bibfnamefont {D.~N.}\ \bibnamefont {Christodoulides}}, \ and\
  \bibinfo {author} {\bibfnamefont {M.}~\bibnamefont {Khajavikhan}},\
  }\href@noop {} {\bibfield  {journal} {\bibinfo  {journal} {Science}\ }\textbf
  {\bibinfo {volume} {346}},\ \bibinfo {pages} {975} (\bibinfo {year}
  {2014})}\BibitemShut {NoStop}%
\bibitem [{\citenamefont {Feng}\ \emph {et~al.}(2014)\citenamefont {Feng},
  \citenamefont {Wong}, \citenamefont {Ma}, \citenamefont {Wang},\ and\
  \citenamefont {Zhang}}]{feng2014single}%
  \BibitemOpen
  \bibfield  {author} {\bibinfo {author} {\bibfnamefont {L.}~\bibnamefont
  {Feng}}, \bibinfo {author} {\bibfnamefont {Z.~J.}\ \bibnamefont {Wong}},
  \bibinfo {author} {\bibfnamefont {R.-M.}\ \bibnamefont {Ma}}, \bibinfo
  {author} {\bibfnamefont {Y.}~\bibnamefont {Wang}}, \ and\ \bibinfo {author}
  {\bibfnamefont {X.}~\bibnamefont {Zhang}},\ }\href@noop {} {\bibfield
  {journal} {\bibinfo  {journal} {Science}\ }\textbf {\bibinfo {volume}
  {346}},\ \bibinfo {pages} {972} (\bibinfo {year} {2014})}\BibitemShut
  {NoStop}%
\bibitem [{\citenamefont {Schumer}\ \emph {et~al.}(2022)\citenamefont
  {Schumer}, \citenamefont {Liu}, \citenamefont {Leshin}, \citenamefont {Ding},
  \citenamefont {Alahmadi}, \citenamefont {Hassan}, \citenamefont {Nasari},
  \citenamefont {Rotter}, \citenamefont {Christodoulides}, \citenamefont
  {LiKamWa},\ and\ \citenamefont {Khajavikhan}}]{schumer2022topological}%
  \BibitemOpen
  \bibfield  {author} {\bibinfo {author} {\bibfnamefont {A.}~\bibnamefont
  {Schumer}}, \bibinfo {author} {\bibfnamefont {Y.}~\bibnamefont {Liu}},
  \bibinfo {author} {\bibfnamefont {J.}~\bibnamefont {Leshin}}, \bibinfo
  {author} {\bibfnamefont {L.}~\bibnamefont {Ding}}, \bibinfo {author}
  {\bibfnamefont {Y.}~\bibnamefont {Alahmadi}}, \bibinfo {author}
  {\bibfnamefont {A.}~\bibnamefont {Hassan}}, \bibinfo {author} {\bibfnamefont
  {H.}~\bibnamefont {Nasari}}, \bibinfo {author} {\bibfnamefont
  {S.}~\bibnamefont {Rotter}}, \bibinfo {author} {\bibfnamefont
  {D.}~\bibnamefont {Christodoulides}}, \bibinfo {author} {\bibfnamefont
  {P.}~\bibnamefont {LiKamWa}}, \ and\ \bibinfo {author} {\bibfnamefont
  {M.}~\bibnamefont {Khajavikhan}},\ }\href@noop {} {\bibfield  {journal}
  {\bibinfo  {journal} {Science}\ }\textbf {\bibinfo {volume} {375}},\ \bibinfo
  {pages} {884} (\bibinfo {year} {2022})}\BibitemShut {NoStop}%
\bibitem [{\citenamefont {Heiss}(2012)}]{heiss2012physics}%
  \BibitemOpen
  \bibfield  {author} {\bibinfo {author} {\bibfnamefont {W.}~\bibnamefont
  {Heiss}},\ }\href@noop {} {\bibfield  {journal} {\bibinfo  {journal} {J.
  Phys. A: Math. Theor.}\ }\textbf {\bibinfo {volume} {45}},\ \bibinfo {pages}
  {444016} (\bibinfo {year} {2012})}\BibitemShut {NoStop}%
\bibitem [{\citenamefont {Zhen}\ \emph {et~al.}(2015)\citenamefont {Zhen},
  \citenamefont {Hsu}, \citenamefont {Igarashi}, \citenamefont {Lu},
  \citenamefont {Kaminer}, \citenamefont {Pick}, \citenamefont {Chua},
  \citenamefont {Joannopoulos},\ and\ \citenamefont
  {Solja{\v{c}}i{\'c}}}]{zhen2015spawning}%
  \BibitemOpen
  \bibfield  {author} {\bibinfo {author} {\bibfnamefont {B.}~\bibnamefont
  {Zhen}}, \bibinfo {author} {\bibfnamefont {C.~W.}\ \bibnamefont {Hsu}},
  \bibinfo {author} {\bibfnamefont {Y.}~\bibnamefont {Igarashi}}, \bibinfo
  {author} {\bibfnamefont {L.}~\bibnamefont {Lu}}, \bibinfo {author}
  {\bibfnamefont {I.}~\bibnamefont {Kaminer}}, \bibinfo {author} {\bibfnamefont
  {A.}~\bibnamefont {Pick}}, \bibinfo {author} {\bibfnamefont {S.-L.}\
  \bibnamefont {Chua}}, \bibinfo {author} {\bibfnamefont {J.~D.}\ \bibnamefont
  {Joannopoulos}}, \ and\ \bibinfo {author} {\bibfnamefont {M.}~\bibnamefont
  {Solja{\v{c}}i{\'c}}},\ }\href@noop {} {\bibfield  {journal} {\bibinfo
  {journal} {Nature}\ }\textbf {\bibinfo {volume} {525}},\ \bibinfo {pages}
  {354} (\bibinfo {year} {2015})}\BibitemShut {NoStop}%
\bibitem [{\citenamefont {Xu}\ \emph {et~al.}(2017)\citenamefont {Xu},
  \citenamefont {Wang},\ and\ \citenamefont {Duan}}]{Xu2017}%
  \BibitemOpen
  \bibfield  {author} {\bibinfo {author} {\bibfnamefont {Y.}~\bibnamefont
  {Xu}}, \bibinfo {author} {\bibfnamefont {S.-T.}\ \bibnamefont {Wang}}, \ and\
  \bibinfo {author} {\bibfnamefont {L.-M.}\ \bibnamefont {Duan}},\ }\href
  {\doibase 10.1103/PhysRevLett.118.045701} {\bibfield  {journal} {\bibinfo
  {journal} {Phys. Rev. Lett.}\ }\textbf {\bibinfo {volume} {118}},\ \bibinfo
  {pages} {045701} (\bibinfo {year} {2017})}\BibitemShut {NoStop}%
\bibitem [{\citenamefont {Cerjan}\ \emph {et~al.}(2019)\citenamefont {Cerjan},
  \citenamefont {Huang}, \citenamefont {Wang}, \citenamefont {Chen},
  \citenamefont {Chong},\ and\ \citenamefont {Rechtsman}}]{cerjan2019}%
  \BibitemOpen
  \bibfield  {author} {\bibinfo {author} {\bibfnamefont {A.}~\bibnamefont
  {Cerjan}}, \bibinfo {author} {\bibfnamefont {S.}~\bibnamefont {Huang}},
  \bibinfo {author} {\bibfnamefont {M.}~\bibnamefont {Wang}}, \bibinfo {author}
  {\bibfnamefont {K.~P.}\ \bibnamefont {Chen}}, \bibinfo {author}
  {\bibfnamefont {Y.}~\bibnamefont {Chong}}, \ and\ \bibinfo {author}
  {\bibfnamefont {M.~C.}\ \bibnamefont {Rechtsman}},\ }\href@noop {} {\bibfield
   {journal} {\bibinfo  {journal} {Nat. Photon.}\ }\textbf {\bibinfo {volume}
  {13}},\ \bibinfo {pages} {623} (\bibinfo {year} {2019})}\BibitemShut
  {NoStop}%
\bibitem [{\citenamefont {Miri}\ and\ \citenamefont
  {Al{\`u}}(2019)}]{miri2019exceptional}%
  \BibitemOpen
  \bibfield  {author} {\bibinfo {author} {\bibfnamefont {M.-A.}\ \bibnamefont
  {Miri}}\ and\ \bibinfo {author} {\bibfnamefont {A.}~\bibnamefont {Al{\`u}}},\
  }\href@noop {} {\bibfield  {journal} {\bibinfo  {journal} {Science}\ }\textbf
  {\bibinfo {volume} {363}},\ \bibinfo {pages} {eaar7709} (\bibinfo {year}
  {2019})}\BibitemShut {NoStop}%
\bibitem [{\citenamefont {Hodaei}\ \emph {et~al.}(2017)\citenamefont {Hodaei},
  \citenamefont {Hassan}, \citenamefont {Wittek}, \citenamefont
  {Garcia-Gracia}, \citenamefont {El-Ganainy}, \citenamefont
  {Christodoulides},\ and\ \citenamefont {Khajavikhan}}]{hodaei2017}%
  \BibitemOpen
  \bibfield  {author} {\bibinfo {author} {\bibfnamefont {H.}~\bibnamefont
  {Hodaei}}, \bibinfo {author} {\bibfnamefont {A.~U.}\ \bibnamefont {Hassan}},
  \bibinfo {author} {\bibfnamefont {S.}~\bibnamefont {Wittek}}, \bibinfo
  {author} {\bibfnamefont {H.}~\bibnamefont {Garcia-Gracia}}, \bibinfo {author}
  {\bibfnamefont {R.}~\bibnamefont {El-Ganainy}}, \bibinfo {author}
  {\bibfnamefont {D.~N.}\ \bibnamefont {Christodoulides}}, \ and\ \bibinfo
  {author} {\bibfnamefont {M.}~\bibnamefont {Khajavikhan}},\ }\href@noop {}
  {\bibfield  {journal} {\bibinfo  {journal} {Nature}\ }\textbf {\bibinfo
  {volume} {548}},\ \bibinfo {pages} {187} (\bibinfo {year}
  {2017})}\BibitemShut {NoStop}%
\bibitem [{\citenamefont {Chen}\ \emph {et~al.}(2017)\citenamefont {Chen},
  \citenamefont {Kaya~{\"O}zdemir}, \citenamefont {Zhao}, \citenamefont
  {Wiersig},\ and\ \citenamefont {Yang}}]{chen2017exceptional}%
  \BibitemOpen
  \bibfield  {author} {\bibinfo {author} {\bibfnamefont {W.}~\bibnamefont
  {Chen}}, \bibinfo {author} {\bibfnamefont {{\c{S}}.}~\bibnamefont
  {Kaya~{\"O}zdemir}}, \bibinfo {author} {\bibfnamefont {G.}~\bibnamefont
  {Zhao}}, \bibinfo {author} {\bibfnamefont {J.}~\bibnamefont {Wiersig}}, \
  and\ \bibinfo {author} {\bibfnamefont {L.}~\bibnamefont {Yang}},\ }\href@noop
  {} {\bibfield  {journal} {\bibinfo  {journal} {Nature}\ }\textbf {\bibinfo
  {volume} {548}},\ \bibinfo {pages} {192} (\bibinfo {year}
  {2017})}\BibitemShut {NoStop}%
\bibitem [{\citenamefont {Hokmabadi}\ \emph {et~al.}(2019)\citenamefont
  {Hokmabadi}, \citenamefont {Schumer}, \citenamefont {Christodoulides},\ and\
  \citenamefont {Khajavikhan}}]{hokmabadi2019non}%
  \BibitemOpen
  \bibfield  {author} {\bibinfo {author} {\bibfnamefont {M.~P.}\ \bibnamefont
  {Hokmabadi}}, \bibinfo {author} {\bibfnamefont {A.}~\bibnamefont {Schumer}},
  \bibinfo {author} {\bibfnamefont {D.~N.}\ \bibnamefont {Christodoulides}}, \
  and\ \bibinfo {author} {\bibfnamefont {M.}~\bibnamefont {Khajavikhan}},\
  }\href@noop {} {\bibfield  {journal} {\bibinfo  {journal} {Nature}\ }\textbf
  {\bibinfo {volume} {576}},\ \bibinfo {pages} {70} (\bibinfo {year}
  {2019})}\BibitemShut {NoStop}%
\bibitem [{\citenamefont {Hatano}\ and\ \citenamefont
  {Nelson}(1996)}]{hatano1996localization}%
  \BibitemOpen
  \bibfield  {author} {\bibinfo {author} {\bibfnamefont {N.}~\bibnamefont
  {Hatano}}\ and\ \bibinfo {author} {\bibfnamefont {D.~R.}\ \bibnamefont
  {Nelson}},\ }\href@noop {} {\bibfield  {journal} {\bibinfo  {journal} {Phys.
  Rev. Lett.}\ }\textbf {\bibinfo {volume} {77}},\ \bibinfo {pages} {570}
  (\bibinfo {year} {1996})}\BibitemShut {NoStop}%
\bibitem [{\citenamefont {Hatano}\ and\ \citenamefont
  {Nelson}(1997)}]{hatano1997vortex}%
  \BibitemOpen
  \bibfield  {author} {\bibinfo {author} {\bibfnamefont {N.}~\bibnamefont
  {Hatano}}\ and\ \bibinfo {author} {\bibfnamefont {D.~R.}\ \bibnamefont
  {Nelson}},\ }\href@noop {} {\bibfield  {journal} {\bibinfo  {journal} {Phys.
  Rev. B}\ }\textbf {\bibinfo {volume} {56}},\ \bibinfo {pages} {8651}
  (\bibinfo {year} {1997})}\BibitemShut {NoStop}%
\bibitem [{\citenamefont {Lee}(2016)}]{lee2016anomalous}%
  \BibitemOpen
  \bibfield  {author} {\bibinfo {author} {\bibfnamefont {T.~E.}\ \bibnamefont
  {Lee}},\ }\href@noop {} {\bibfield  {journal} {\bibinfo  {journal} {Phys.
  Rev. Lett.}\ }\textbf {\bibinfo {volume} {116}},\ \bibinfo {pages} {133903}
  (\bibinfo {year} {2016})}\BibitemShut {NoStop}%
\bibitem [{\citenamefont {Kunst}\ \emph {et~al.}(2018)\citenamefont {Kunst},
  \citenamefont {Edvardsson}, \citenamefont {Budich},\ and\ \citenamefont
  {Bergholtz}}]{kunst2018biorthogonal}%
  \BibitemOpen
  \bibfield  {author} {\bibinfo {author} {\bibfnamefont {F.~K.}\ \bibnamefont
  {Kunst}}, \bibinfo {author} {\bibfnamefont {E.}~\bibnamefont {Edvardsson}},
  \bibinfo {author} {\bibfnamefont {J.~C.}\ \bibnamefont {Budich}}, \ and\
  \bibinfo {author} {\bibfnamefont {E.~J.}\ \bibnamefont {Bergholtz}},\
  }\href@noop {} {\bibfield  {journal} {\bibinfo  {journal} {Phys. Rev. Lett.}\
  }\textbf {\bibinfo {volume} {121}},\ \bibinfo {pages} {026808} (\bibinfo
  {year} {2018})}\BibitemShut {NoStop}%
\bibitem [{\citenamefont {Yao}\ and\ \citenamefont {Wang}(2018)}]{yao2018edge}%
  \BibitemOpen
  \bibfield  {author} {\bibinfo {author} {\bibfnamefont {S.}~\bibnamefont
  {Yao}}\ and\ \bibinfo {author} {\bibfnamefont {Z.}~\bibnamefont {Wang}},\
  }\href@noop {} {\bibfield  {journal} {\bibinfo  {journal} {Phys. Rev. Lett.}\
  }\textbf {\bibinfo {volume} {121}},\ \bibinfo {pages} {086803} (\bibinfo
  {year} {2018})}\BibitemShut {NoStop}%
\bibitem [{\citenamefont {Yokomizo}\ and\ \citenamefont
  {Murakami}(2019)}]{yokomizo2019non}%
  \BibitemOpen
  \bibfield  {author} {\bibinfo {author} {\bibfnamefont {K.}~\bibnamefont
  {Yokomizo}}\ and\ \bibinfo {author} {\bibfnamefont {S.}~\bibnamefont
  {Murakami}},\ }\href@noop {} {\bibfield  {journal} {\bibinfo  {journal}
  {Phys. Rev. Lett.}\ }\textbf {\bibinfo {volume} {123}},\ \bibinfo {pages}
  {066404} (\bibinfo {year} {2019})}\BibitemShut {NoStop}%
\bibitem [{\citenamefont {Borgnia}\ \emph {et~al.}(2020)\citenamefont
  {Borgnia}, \citenamefont {Kruchkov},\ and\ \citenamefont
  {Slager}}]{borgnia2020non}%
  \BibitemOpen
  \bibfield  {author} {\bibinfo {author} {\bibfnamefont {D.~S.}\ \bibnamefont
  {Borgnia}}, \bibinfo {author} {\bibfnamefont {A.~J.}\ \bibnamefont
  {Kruchkov}}, \ and\ \bibinfo {author} {\bibfnamefont {R.-J.}\ \bibnamefont
  {Slager}},\ }\href@noop {} {\bibfield  {journal} {\bibinfo  {journal} {Phys.
  Rev. Lett.}\ }\textbf {\bibinfo {volume} {124}},\ \bibinfo {pages} {056802}
  (\bibinfo {year} {2020})}\BibitemShut {NoStop}%
\bibitem [{\citenamefont {Okuma}\ \emph {et~al.}(2020)\citenamefont {Okuma},
  \citenamefont {Kawabata}, \citenamefont {Shiozaki},\ and\ \citenamefont
  {Sato}}]{okuma2020topological}%
  \BibitemOpen
  \bibfield  {author} {\bibinfo {author} {\bibfnamefont {N.}~\bibnamefont
  {Okuma}}, \bibinfo {author} {\bibfnamefont {K.}~\bibnamefont {Kawabata}},
  \bibinfo {author} {\bibfnamefont {K.}~\bibnamefont {Shiozaki}}, \ and\
  \bibinfo {author} {\bibfnamefont {M.}~\bibnamefont {Sato}},\ }\href@noop {}
  {\bibfield  {journal} {\bibinfo  {journal} {Phys. Rev. Lett.}\ }\textbf
  {\bibinfo {volume} {124}},\ \bibinfo {pages} {086801} (\bibinfo {year}
  {2020})}\BibitemShut {NoStop}%
\bibitem [{\citenamefont {Zhang}\ \emph {et~al.}(2020)\citenamefont {Zhang},
  \citenamefont {Yang},\ and\ \citenamefont {Fang}}]{zhang2020correspondence}%
  \BibitemOpen
  \bibfield  {author} {\bibinfo {author} {\bibfnamefont {K.}~\bibnamefont
  {Zhang}}, \bibinfo {author} {\bibfnamefont {Z.}~\bibnamefont {Yang}}, \ and\
  \bibinfo {author} {\bibfnamefont {C.}~\bibnamefont {Fang}},\ }\href@noop {}
  {\bibfield  {journal} {\bibinfo  {journal} {Phys. Rev. Lett.}\ }\textbf
  {\bibinfo {volume} {125}},\ \bibinfo {pages} {126402} (\bibinfo {year}
  {2020})}\BibitemShut {NoStop}%
\bibitem [{\citenamefont {Zhang}\ \emph {et~al.}(2022)\citenamefont {Zhang},
  \citenamefont {Yang},\ and\ \citenamefont {Fang}}]{zhang2022universal}%
  \BibitemOpen
  \bibfield  {author} {\bibinfo {author} {\bibfnamefont {K.}~\bibnamefont
  {Zhang}}, \bibinfo {author} {\bibfnamefont {Z.}~\bibnamefont {Yang}}, \ and\
  \bibinfo {author} {\bibfnamefont {C.}~\bibnamefont {Fang}},\ }\href@noop {}
  {\bibfield  {journal} {\bibinfo  {journal} {Nat. Comm.}\ }\textbf {\bibinfo
  {volume} {13}},\ \bibinfo {pages} {2496} (\bibinfo {year}
  {2022})}\BibitemShut {NoStop}%
\bibitem [{\citenamefont {Weidemann}\ \emph {et~al.}(2020)\citenamefont
  {Weidemann}, \citenamefont {Kremer}, \citenamefont {Helbig}, \citenamefont
  {Hofmann}, \citenamefont {Stegmaier}, \citenamefont {Greiter}, \citenamefont
  {Thomale},\ and\ \citenamefont {Szameit}}]{weidemann2020topological}%
  \BibitemOpen
  \bibfield  {author} {\bibinfo {author} {\bibfnamefont {S.}~\bibnamefont
  {Weidemann}}, \bibinfo {author} {\bibfnamefont {M.}~\bibnamefont {Kremer}},
  \bibinfo {author} {\bibfnamefont {T.}~\bibnamefont {Helbig}}, \bibinfo
  {author} {\bibfnamefont {T.}~\bibnamefont {Hofmann}}, \bibinfo {author}
  {\bibfnamefont {A.}~\bibnamefont {Stegmaier}}, \bibinfo {author}
  {\bibfnamefont {M.}~\bibnamefont {Greiter}}, \bibinfo {author} {\bibfnamefont
  {R.}~\bibnamefont {Thomale}}, \ and\ \bibinfo {author} {\bibfnamefont
  {A.}~\bibnamefont {Szameit}},\ }\href@noop {} {\bibfield  {journal} {\bibinfo
   {journal} {Science}\ }\textbf {\bibinfo {volume} {368}},\ \bibinfo {pages}
  {311} (\bibinfo {year} {2020})}\BibitemShut {NoStop}%
\bibitem [{\citenamefont {Longhi}()}]{Longhi2018}%
  \BibitemOpen
  \bibfield  {author} {\bibinfo {author} {\bibfnamefont {S.}~\bibnamefont
  {Longhi}},\ }\href@noop {} {\bibfield  {journal} {\bibinfo  {journal} {Ann.
  Phys.}\ }\textbf {\bibinfo {volume} {530}},\ \bibinfo {pages}
  {1800023}}\BibitemShut {NoStop}%
\bibitem [{\citenamefont {Zhu}\ \emph {et~al.}(2022)\citenamefont {Zhu},
  \citenamefont {Wang}, \citenamefont {Leykam}, \citenamefont {Xue},
  \citenamefont {Wang},\ and\ \citenamefont {Chong}}]{zhu2022anomalous}%
  \BibitemOpen
  \bibfield  {author} {\bibinfo {author} {\bibfnamefont {B.}~\bibnamefont
  {Zhu}}, \bibinfo {author} {\bibfnamefont {Q.}~\bibnamefont {Wang}}, \bibinfo
  {author} {\bibfnamefont {D.}~\bibnamefont {Leykam}}, \bibinfo {author}
  {\bibfnamefont {H.}~\bibnamefont {Xue}}, \bibinfo {author} {\bibfnamefont
  {Q.~J.}\ \bibnamefont {Wang}}, \ and\ \bibinfo {author} {\bibfnamefont
  {Y.}~\bibnamefont {Chong}},\ }\href@noop {} {\bibfield  {journal} {\bibinfo
  {journal} {Phys. Rev. Lett.}\ }\textbf {\bibinfo {volume} {129}},\ \bibinfo
  {pages} {013903} (\bibinfo {year} {2022})}\BibitemShut {NoStop}%
\bibitem [{SM(2022)}]{SM}%
  \BibitemOpen
  \href@noop {} {\enquote {\bibinfo {title} {See supplementary material},}\ }
  (\bibinfo {year} {2022}),\ \bibinfo {note} {see online Supplemental
  Materials.}\BibitemShut {Stop}%
\bibitem [{\citenamefont {Peres}\ \emph {et~al.}(2006)\citenamefont {Peres},
  \citenamefont {Guinea},\ and\ \citenamefont {Castro~Neto}}]{Peres2006}%
  \BibitemOpen
  \bibfield  {author} {\bibinfo {author} {\bibfnamefont {N.~M.~R.}\
  \bibnamefont {Peres}}, \bibinfo {author} {\bibfnamefont {F.}~\bibnamefont
  {Guinea}}, \ and\ \bibinfo {author} {\bibfnamefont {A.~H.}\ \bibnamefont
  {Castro~Neto}},\ }\href {\doibase 10.1103/PhysRevB.73.125411} {\bibfield
  {journal} {\bibinfo  {journal} {Phys. Rev. B}\ }\textbf {\bibinfo {volume}
  {73}},\ \bibinfo {pages} {125411} (\bibinfo {year} {2006})}\BibitemShut
  {NoStop}%
\bibitem [{\citenamefont {Zhang}\ \emph
  {et~al.}(2006{\natexlab{b}})\citenamefont {Zhang}, \citenamefont {Jiang},
  \citenamefont {Small}, \citenamefont {Purewal}, \citenamefont {Tan},
  \citenamefont {Fazlollahi}, \citenamefont {Chudow}, \citenamefont {Jaszczak},
  \citenamefont {Stormer},\ and\ \citenamefont {Kim}}]{ZhangGraphene2006}%
  \BibitemOpen
  \bibfield  {author} {\bibinfo {author} {\bibfnamefont {Y.}~\bibnamefont
  {Zhang}}, \bibinfo {author} {\bibfnamefont {Z.}~\bibnamefont {Jiang}},
  \bibinfo {author} {\bibfnamefont {J.~P.}\ \bibnamefont {Small}}, \bibinfo
  {author} {\bibfnamefont {M.~S.}\ \bibnamefont {Purewal}}, \bibinfo {author}
  {\bibfnamefont {Y.-W.}\ \bibnamefont {Tan}}, \bibinfo {author} {\bibfnamefont
  {M.}~\bibnamefont {Fazlollahi}}, \bibinfo {author} {\bibfnamefont {J.~D.}\
  \bibnamefont {Chudow}}, \bibinfo {author} {\bibfnamefont {J.~A.}\
  \bibnamefont {Jaszczak}}, \bibinfo {author} {\bibfnamefont {H.~L.}\
  \bibnamefont {Stormer}}, \ and\ \bibinfo {author} {\bibfnamefont
  {P.}~\bibnamefont {Kim}},\ }\href {\doibase 10.1103/PhysRevLett.96.136806}
  {\bibfield  {journal} {\bibinfo  {journal} {Phys. Rev. Lett.}\ }\textbf
  {\bibinfo {volume} {96}},\ \bibinfo {pages} {136806} (\bibinfo {year}
  {2006}{\natexlab{b}})}\BibitemShut {NoStop}%
\bibitem [{\citenamefont {Nomura}\ and\ \citenamefont
  {MacDonald}(2006)}]{Kentaro2006}%
  \BibitemOpen
  \bibfield  {author} {\bibinfo {author} {\bibfnamefont {K.}~\bibnamefont
  {Nomura}}\ and\ \bibinfo {author} {\bibfnamefont {A.~H.}\ \bibnamefont
  {MacDonald}},\ }\href {\doibase 10.1103/PhysRevLett.96.256602} {\bibfield
  {journal} {\bibinfo  {journal} {Phys. Rev. Lett.}\ }\textbf {\bibinfo
  {volume} {96}},\ \bibinfo {pages} {256602} (\bibinfo {year}
  {2006})}\BibitemShut {NoStop}%
\bibitem [{\citenamefont {Guinea}\ \emph {et~al.}(2010)\citenamefont {Guinea},
  \citenamefont {Katsnelson},\ and\ \citenamefont {Geim}}]{guinea2010energy}%
  \BibitemOpen
  \bibfield  {author} {\bibinfo {author} {\bibfnamefont {F.}~\bibnamefont
  {Guinea}}, \bibinfo {author} {\bibfnamefont {M.}~\bibnamefont {Katsnelson}},
  \ and\ \bibinfo {author} {\bibfnamefont {A.}~\bibnamefont {Geim}},\
  }\href@noop {} {\bibfield  {journal} {\bibinfo  {journal} {Nat. Phys.}\
  }\textbf {\bibinfo {volume} {6}},\ \bibinfo {pages} {30} (\bibinfo {year}
  {2010})}\BibitemShut {NoStop}%
\bibitem [{\citenamefont {Rechtsman}\ \emph {et~al.}(2013)\citenamefont
  {Rechtsman}, \citenamefont {Zeuner}, \citenamefont {T{\"u}nnermann},
  \citenamefont {Nolte}, \citenamefont {Segev},\ and\ \citenamefont
  {Szameit}}]{rechtsman2013strain}%
  \BibitemOpen
  \bibfield  {author} {\bibinfo {author} {\bibfnamefont {M.~C.}\ \bibnamefont
  {Rechtsman}}, \bibinfo {author} {\bibfnamefont {J.~M.}\ \bibnamefont
  {Zeuner}}, \bibinfo {author} {\bibfnamefont {A.}~\bibnamefont
  {T{\"u}nnermann}}, \bibinfo {author} {\bibfnamefont {S.}~\bibnamefont
  {Nolte}}, \bibinfo {author} {\bibfnamefont {M.}~\bibnamefont {Segev}}, \ and\
  \bibinfo {author} {\bibfnamefont {A.}~\bibnamefont {Szameit}},\ }\href@noop
  {} {\bibfield  {journal} {\bibinfo  {journal} {Nat. Photon.}\ }\textbf
  {\bibinfo {volume} {7}},\ \bibinfo {pages} {153} (\bibinfo {year}
  {2013})}\BibitemShut {NoStop}%
\bibitem [{\citenamefont {Schomerus}\ and\ \citenamefont
  {Halpern}(2013)}]{schomerus2013parity}%
  \BibitemOpen
  \bibfield  {author} {\bibinfo {author} {\bibfnamefont {H.}~\bibnamefont
  {Schomerus}}\ and\ \bibinfo {author} {\bibfnamefont {N.~Y.}\ \bibnamefont
  {Halpern}},\ }\href@noop {} {\bibfield  {journal} {\bibinfo  {journal} {Phys.
  Rev. Lett.}\ }\textbf {\bibinfo {volume} {110}},\ \bibinfo {pages} {013903}
  (\bibinfo {year} {2013})}\BibitemShut {NoStop}%
\bibitem [{\citenamefont {Shen}\ and\ \citenamefont
  {Fu}(2018)}]{shen2018quantum}%
  \BibitemOpen
  \bibfield  {author} {\bibinfo {author} {\bibfnamefont {H.}~\bibnamefont
  {Shen}}\ and\ \bibinfo {author} {\bibfnamefont {L.}~\bibnamefont {Fu}},\
  }\href@noop {} {\bibfield  {journal} {\bibinfo  {journal} {Phys. Rev. Lett.}\
  }\textbf {\bibinfo {volume} {121}},\ \bibinfo {pages} {026403} (\bibinfo
  {year} {2018})}\BibitemShut {NoStop}%
\bibitem [{\citenamefont {Lu}\ \emph {et~al.}(2021{\natexlab{a}})\citenamefont
  {Lu}, \citenamefont {Zhang},\ and\ \citenamefont {Franz}}]{lu2021magnetic}%
  \BibitemOpen
  \bibfield  {author} {\bibinfo {author} {\bibfnamefont {M.}~\bibnamefont
  {Lu}}, \bibinfo {author} {\bibfnamefont {X.-X.}\ \bibnamefont {Zhang}}, \
  and\ \bibinfo {author} {\bibfnamefont {M.}~\bibnamefont {Franz}},\
  }\href@noop {} {\bibfield  {journal} {\bibinfo  {journal} {Phys. Rev. Lett.}\
  }\textbf {\bibinfo {volume} {127}},\ \bibinfo {pages} {256402} (\bibinfo
  {year} {2021}{\natexlab{a}})}\BibitemShut {NoStop}%
\bibitem [{\citenamefont {Shao}\ \emph {et~al.}(2022)\citenamefont {Shao},
  \citenamefont {Cai}, \citenamefont {Geng}, \citenamefont {Chen},\ and\
  \citenamefont {Xing}}]{shao2022cyclotron}%
  \BibitemOpen
  \bibfield  {author} {\bibinfo {author} {\bibfnamefont {K.}~\bibnamefont
  {Shao}}, \bibinfo {author} {\bibfnamefont {Z.-T.}\ \bibnamefont {Cai}},
  \bibinfo {author} {\bibfnamefont {H.}~\bibnamefont {Geng}}, \bibinfo {author}
  {\bibfnamefont {W.}~\bibnamefont {Chen}}, \ and\ \bibinfo {author}
  {\bibfnamefont {D.}~\bibnamefont {Xing}},\ }\href@noop {} {\bibfield
  {journal} {\bibinfo  {journal} {Phys. Rev. B}\ }\textbf {\bibinfo {volume}
  {106}},\ \bibinfo {pages} {L081402} (\bibinfo {year} {2022})}\BibitemShut
  {NoStop}%
\bibitem [{\citenamefont {Zhu}\ \emph {et~al.}(2020)\citenamefont {Zhu},
  \citenamefont {Wang}, \citenamefont {Gupta}, \citenamefont {Zhang},
  \citenamefont {Xie}, \citenamefont {Lu},\ and\ \citenamefont
  {Chen}}]{zhu2020photonic}%
  \BibitemOpen
  \bibfield  {author} {\bibinfo {author} {\bibfnamefont {X.}~\bibnamefont
  {Zhu}}, \bibinfo {author} {\bibfnamefont {H.}~\bibnamefont {Wang}}, \bibinfo
  {author} {\bibfnamefont {S.~K.}\ \bibnamefont {Gupta}}, \bibinfo {author}
  {\bibfnamefont {H.}~\bibnamefont {Zhang}}, \bibinfo {author} {\bibfnamefont
  {B.}~\bibnamefont {Xie}}, \bibinfo {author} {\bibfnamefont {M.}~\bibnamefont
  {Lu}}, \ and\ \bibinfo {author} {\bibfnamefont {Y.}~\bibnamefont {Chen}},\
  }\href@noop {} {\bibfield  {journal} {\bibinfo  {journal} {Phys. Rev. Res.}\
  }\textbf {\bibinfo {volume} {2}},\ \bibinfo {pages} {013280} (\bibinfo {year}
  {2020})}\BibitemShut {NoStop}%
\bibitem [{\citenamefont {Song}\ \emph {et~al.}(2020)\citenamefont {Song},
  \citenamefont {Liu}, \citenamefont {Zheng}, \citenamefont {Zhang},
  \citenamefont {Wang},\ and\ \citenamefont {Lu}}]{song2020two}%
  \BibitemOpen
  \bibfield  {author} {\bibinfo {author} {\bibfnamefont {Y.}~\bibnamefont
  {Song}}, \bibinfo {author} {\bibfnamefont {W.}~\bibnamefont {Liu}}, \bibinfo
  {author} {\bibfnamefont {L.}~\bibnamefont {Zheng}}, \bibinfo {author}
  {\bibfnamefont {Y.}~\bibnamefont {Zhang}}, \bibinfo {author} {\bibfnamefont
  {B.}~\bibnamefont {Wang}}, \ and\ \bibinfo {author} {\bibfnamefont
  {P.}~\bibnamefont {Lu}},\ }\href@noop {} {\bibfield  {journal} {\bibinfo
  {journal} {Phys. Rev. Appl.}\ }\textbf {\bibinfo {volume} {14}},\ \bibinfo
  {pages} {064076} (\bibinfo {year} {2020})}\BibitemShut {NoStop}%
\bibitem [{\citenamefont {Helbig}\ \emph {et~al.}(2020)\citenamefont {Helbig},
  \citenamefont {Hofmann}, \citenamefont {Imhof}, \citenamefont {Abdelghany},
  \citenamefont {Kiessling}, \citenamefont {Molenkamp}, \citenamefont {Lee},
  \citenamefont {Szameit}, \citenamefont {Greiter},\ and\ \citenamefont
  {Thomale}}]{helbig2020generalized}%
  \BibitemOpen
  \bibfield  {author} {\bibinfo {author} {\bibfnamefont {T.}~\bibnamefont
  {Helbig}}, \bibinfo {author} {\bibfnamefont {T.}~\bibnamefont {Hofmann}},
  \bibinfo {author} {\bibfnamefont {S.}~\bibnamefont {Imhof}}, \bibinfo
  {author} {\bibfnamefont {M.}~\bibnamefont {Abdelghany}}, \bibinfo {author}
  {\bibfnamefont {T.}~\bibnamefont {Kiessling}}, \bibinfo {author}
  {\bibfnamefont {L.}~\bibnamefont {Molenkamp}}, \bibinfo {author}
  {\bibfnamefont {C.}~\bibnamefont {Lee}}, \bibinfo {author} {\bibfnamefont
  {A.}~\bibnamefont {Szameit}}, \bibinfo {author} {\bibfnamefont
  {M.}~\bibnamefont {Greiter}}, \ and\ \bibinfo {author} {\bibfnamefont
  {R.}~\bibnamefont {Thomale}},\ }\href@noop {} {\bibfield  {journal} {\bibinfo
   {journal} {Nat. Phys.}\ }\textbf {\bibinfo {volume} {16}},\ \bibinfo {pages}
  {747} (\bibinfo {year} {2020})}\BibitemShut {NoStop}%
\bibitem [{\citenamefont {Zou}\ \emph {et~al.}(2021)\citenamefont {Zou},
  \citenamefont {Chen}, \citenamefont {He}, \citenamefont {Bao}, \citenamefont
  {Lee}, \citenamefont {Sun},\ and\ \citenamefont
  {Zhang}}]{zou2021observation}%
  \BibitemOpen
  \bibfield  {author} {\bibinfo {author} {\bibfnamefont {D.}~\bibnamefont
  {Zou}}, \bibinfo {author} {\bibfnamefont {T.}~\bibnamefont {Chen}}, \bibinfo
  {author} {\bibfnamefont {W.}~\bibnamefont {He}}, \bibinfo {author}
  {\bibfnamefont {J.}~\bibnamefont {Bao}}, \bibinfo {author} {\bibfnamefont
  {C.~H.}\ \bibnamefont {Lee}}, \bibinfo {author} {\bibfnamefont
  {H.}~\bibnamefont {Sun}}, \ and\ \bibinfo {author} {\bibfnamefont
  {X.}~\bibnamefont {Zhang}},\ }\href@noop {} {\bibfield  {journal} {\bibinfo
  {journal} {Nat. Comm.}\ }\textbf {\bibinfo {volume} {12}},\ \bibinfo {pages}
  {7201} (\bibinfo {year} {2021})}\BibitemShut {NoStop}%
\bibitem [{\citenamefont {Zhang}\ \emph
  {et~al.}(2021{\natexlab{a}})\citenamefont {Zhang}, \citenamefont {Tian},
  \citenamefont {Jiang}, \citenamefont {Lu},\ and\ \citenamefont
  {Chen}}]{zhang2021observation}%
  \BibitemOpen
  \bibfield  {author} {\bibinfo {author} {\bibfnamefont {X.}~\bibnamefont
  {Zhang}}, \bibinfo {author} {\bibfnamefont {Y.}~\bibnamefont {Tian}},
  \bibinfo {author} {\bibfnamefont {J.-H.}\ \bibnamefont {Jiang}}, \bibinfo
  {author} {\bibfnamefont {M.-H.}\ \bibnamefont {Lu}}, \ and\ \bibinfo {author}
  {\bibfnamefont {Y.-F.}\ \bibnamefont {Chen}},\ }\href@noop {} {\bibfield
  {journal} {\bibinfo  {journal} {Nat. Comm.}\ }\textbf {\bibinfo {volume}
  {12}},\ \bibinfo {pages} {5377} (\bibinfo {year}
  {2021}{\natexlab{a}})}\BibitemShut {NoStop}%
\bibitem [{\citenamefont {Zhang}\ \emph
  {et~al.}(2021{\natexlab{b}})\citenamefont {Zhang}, \citenamefont {Yang},
  \citenamefont {Ge}, \citenamefont {Guan}, \citenamefont {Chen}, \citenamefont
  {Yan}, \citenamefont {Chen}, \citenamefont {Xi}, \citenamefont {Li},
  \citenamefont {Jia}, \citenamefont {Yuan}, \citenamefont {Sun}, \citenamefont
  {Hongsheng},\ and\ \citenamefont {Zhang}}]{zhang2021acoustic}%
  \BibitemOpen
  \bibfield  {author} {\bibinfo {author} {\bibfnamefont {L.}~\bibnamefont
  {Zhang}}, \bibinfo {author} {\bibfnamefont {Y.}~\bibnamefont {Yang}},
  \bibinfo {author} {\bibfnamefont {Y.}~\bibnamefont {Ge}}, \bibinfo {author}
  {\bibfnamefont {Y.-J.}\ \bibnamefont {Guan}}, \bibinfo {author}
  {\bibfnamefont {Q.}~\bibnamefont {Chen}}, \bibinfo {author} {\bibfnamefont
  {Q.}~\bibnamefont {Yan}}, \bibinfo {author} {\bibfnamefont {F.}~\bibnamefont
  {Chen}}, \bibinfo {author} {\bibfnamefont {R.}~\bibnamefont {Xi}}, \bibinfo
  {author} {\bibfnamefont {Y.}~\bibnamefont {Li}}, \bibinfo {author}
  {\bibfnamefont {D.}~\bibnamefont {Jia}}, \bibinfo {author} {\bibfnamefont
  {S.-Q.}\ \bibnamefont {Yuan}}, \bibinfo {author} {\bibfnamefont {H.-X.}\
  \bibnamefont {Sun}}, \bibinfo {author} {\bibfnamefont {C.}~\bibnamefont
  {Hongsheng}}, \ and\ \bibinfo {author} {\bibfnamefont {B.}~\bibnamefont
  {Zhang}},\ }\href@noop {} {\bibfield  {journal} {\bibinfo  {journal} {Nat.
  Comm.}\ }\textbf {\bibinfo {volume} {12}},\ \bibinfo {pages} {6297} (\bibinfo
  {year} {2021}{\natexlab{b}})}\BibitemShut {NoStop}%
\bibitem [{\citenamefont {Gao}\ \emph {et~al.}(2022)\citenamefont {Gao},
  \citenamefont {Xue}, \citenamefont {Gu}, \citenamefont {Li}, \citenamefont
  {Zhu}, \citenamefont {Su}, \citenamefont {Zhu}, \citenamefont {Zhang},\ and\
  \citenamefont {Chong}}]{gao2022non}%
  \BibitemOpen
  \bibfield  {author} {\bibinfo {author} {\bibfnamefont {H.}~\bibnamefont
  {Gao}}, \bibinfo {author} {\bibfnamefont {H.}~\bibnamefont {Xue}}, \bibinfo
  {author} {\bibfnamefont {Z.}~\bibnamefont {Gu}}, \bibinfo {author}
  {\bibfnamefont {L.}~\bibnamefont {Li}}, \bibinfo {author} {\bibfnamefont
  {W.}~\bibnamefont {Zhu}}, \bibinfo {author} {\bibfnamefont {Z.}~\bibnamefont
  {Su}}, \bibinfo {author} {\bibfnamefont {J.}~\bibnamefont {Zhu}}, \bibinfo
  {author} {\bibfnamefont {B.}~\bibnamefont {Zhang}}, \ and\ \bibinfo {author}
  {\bibfnamefont {Y.}~\bibnamefont {Chong}},\ }\href@noop {} {\bibfield
  {journal} {\bibinfo  {journal} {arXiv preprint arXiv:2205.14824}\ } (\bibinfo
  {year} {2022})}\BibitemShut {NoStop}%
\bibitem [{\citenamefont {Liang}\ \emph {et~al.}(2022)\citenamefont {Liang},
  \citenamefont {Xie}, \citenamefont {Dong}, \citenamefont {Li}, \citenamefont
  {Li}, \citenamefont {Gadway}, \citenamefont {Yi},\ and\ \citenamefont
  {Yan}}]{liang2022dynamic}%
  \BibitemOpen
  \bibfield  {author} {\bibinfo {author} {\bibfnamefont {Q.}~\bibnamefont
  {Liang}}, \bibinfo {author} {\bibfnamefont {D.}~\bibnamefont {Xie}}, \bibinfo
  {author} {\bibfnamefont {Z.}~\bibnamefont {Dong}}, \bibinfo {author}
  {\bibfnamefont {H.}~\bibnamefont {Li}}, \bibinfo {author} {\bibfnamefont
  {H.}~\bibnamefont {Li}}, \bibinfo {author} {\bibfnamefont {B.}~\bibnamefont
  {Gadway}}, \bibinfo {author} {\bibfnamefont {W.}~\bibnamefont {Yi}}, \ and\
  \bibinfo {author} {\bibfnamefont {B.}~\bibnamefont {Yan}},\ }\href@noop {}
  {\bibfield  {journal} {\bibinfo  {journal} {Phys. Rev. Lett.}\ }\textbf
  {\bibinfo {volume} {129}},\ \bibinfo {pages} {070401} (\bibinfo {year}
  {2022})}\BibitemShut {NoStop}%
\bibitem [{\citenamefont {Wang}\ \emph {et~al.}(2022)\citenamefont {Wang},
  \citenamefont {Wang},\ and\ \citenamefont {Ma}}]{Wang2022}%
  \BibitemOpen
  \bibfield  {author} {\bibinfo {author} {\bibfnamefont {W.}~\bibnamefont
  {Wang}}, \bibinfo {author} {\bibfnamefont {X.}~\bibnamefont {Wang}}, \ and\
  \bibinfo {author} {\bibfnamefont {G.}~\bibnamefont {Ma}},\ }\href {\doibase
  10.1038/s41586-022-04929-1} {\bibfield  {journal} {\bibinfo  {journal}
  {Nature}\ }\textbf {\bibinfo {volume} {608}},\ \bibinfo {pages} {50}
  (\bibinfo {year} {2022})}\BibitemShut {NoStop}%
\bibitem [{\citenamefont {Tsakmakidis}\ \emph {et~al.}(2007)\citenamefont
  {Tsakmakidis}, \citenamefont {Boardman},\ and\ \citenamefont
  {Hess}}]{tsakmakidis2007trapped}%
  \BibitemOpen
  \bibfield  {author} {\bibinfo {author} {\bibfnamefont {K.~L.}\ \bibnamefont
  {Tsakmakidis}}, \bibinfo {author} {\bibfnamefont {A.~D.}\ \bibnamefont
  {Boardman}}, \ and\ \bibinfo {author} {\bibfnamefont {O.}~\bibnamefont
  {Hess}},\ }\href@noop {} {\bibfield  {journal} {\bibinfo  {journal} {Nature}\
  }\textbf {\bibinfo {volume} {450}},\ \bibinfo {pages} {397} (\bibinfo {year}
  {2007})}\BibitemShut {NoStop}%
\bibitem [{\citenamefont {Gan}\ \emph {et~al.}(2009)\citenamefont {Gan},
  \citenamefont {Ding},\ and\ \citenamefont {Bartoli}}]{gan2009rainbow}%
  \BibitemOpen
  \bibfield  {author} {\bibinfo {author} {\bibfnamefont {Q.}~\bibnamefont
  {Gan}}, \bibinfo {author} {\bibfnamefont {Y.~J.}\ \bibnamefont {Ding}}, \
  and\ \bibinfo {author} {\bibfnamefont {F.~J.}\ \bibnamefont {Bartoli}},\
  }\href@noop {} {\bibfield  {journal} {\bibinfo  {journal} {Physical Review
  Letters}\ }\textbf {\bibinfo {volume} {102}},\ \bibinfo {pages} {056801}
  (\bibinfo {year} {2009})}\BibitemShut {NoStop}%
\bibitem [{\citenamefont {Lu}\ \emph {et~al.}(2021{\natexlab{b}})\citenamefont
  {Lu}, \citenamefont {Wang}, \citenamefont {Xiao}, \citenamefont {Zhang},\
  and\ \citenamefont {Chan}}]{lu2021topological}%
  \BibitemOpen
  \bibfield  {author} {\bibinfo {author} {\bibfnamefont {C.}~\bibnamefont
  {Lu}}, \bibinfo {author} {\bibfnamefont {C.}~\bibnamefont {Wang}}, \bibinfo
  {author} {\bibfnamefont {M.}~\bibnamefont {Xiao}}, \bibinfo {author}
  {\bibfnamefont {Z.}~\bibnamefont {Zhang}}, \ and\ \bibinfo {author}
  {\bibfnamefont {C.~T.}\ \bibnamefont {Chan}},\ }\href@noop {} {\bibfield
  {journal} {\bibinfo  {journal} {Phys. Rev. Lett.}\ }\textbf {\bibinfo
  {volume} {126}},\ \bibinfo {pages} {113902} (\bibinfo {year}
  {2021}{\natexlab{b}})}\BibitemShut {NoStop}%
\bibitem [{\citenamefont {Lu}\ \emph {et~al.}(2022)\citenamefont {Lu},
  \citenamefont {Sun}, \citenamefont {Wang}, \citenamefont {Zhang},
  \citenamefont {Zhao}, \citenamefont {Hu}, \citenamefont {Xiao}, \citenamefont
  {Ding}, \citenamefont {Liu},\ and\ \citenamefont {Chan}}]{lu2022chip}%
  \BibitemOpen
  \bibfield  {author} {\bibinfo {author} {\bibfnamefont {C.}~\bibnamefont
  {Lu}}, \bibinfo {author} {\bibfnamefont {Y.-Z.}\ \bibnamefont {Sun}},
  \bibinfo {author} {\bibfnamefont {C.}~\bibnamefont {Wang}}, \bibinfo {author}
  {\bibfnamefont {H.}~\bibnamefont {Zhang}}, \bibinfo {author} {\bibfnamefont
  {W.}~\bibnamefont {Zhao}}, \bibinfo {author} {\bibfnamefont {X.}~\bibnamefont
  {Hu}}, \bibinfo {author} {\bibfnamefont {M.}~\bibnamefont {Xiao}}, \bibinfo
  {author} {\bibfnamefont {W.}~\bibnamefont {Ding}}, \bibinfo {author}
  {\bibfnamefont {Y.-C.}\ \bibnamefont {Liu}}, \ and\ \bibinfo {author}
  {\bibfnamefont {C.}~\bibnamefont {Chan}},\ }\href@noop {} {\bibfield
  {journal} {\bibinfo  {journal} {Nat. Comm.}\ }\textbf {\bibinfo {volume}
  {13}},\ \bibinfo {pages} {2586} (\bibinfo {year} {2022})}\BibitemShut
  {NoStop}%
\bibitem [{\citenamefont {Peng}\ \emph {et~al.}(2014)\citenamefont {Peng},
  \citenamefont {{\"O}zdemir}, \citenamefont {Lei}, \citenamefont {Monifi},
  \citenamefont {Gianfreda}, \citenamefont {Long}, \citenamefont {Fan},
  \citenamefont {Nori}, \citenamefont {Bender},\ and\ \citenamefont
  {Yang}}]{peng2014parity}%
  \BibitemOpen
  \bibfield  {author} {\bibinfo {author} {\bibfnamefont {B.}~\bibnamefont
  {Peng}}, \bibinfo {author} {\bibfnamefont {{\c{S}}.~K.}\ \bibnamefont
  {{\"O}zdemir}}, \bibinfo {author} {\bibfnamefont {F.}~\bibnamefont {Lei}},
  \bibinfo {author} {\bibfnamefont {F.}~\bibnamefont {Monifi}}, \bibinfo
  {author} {\bibfnamefont {M.}~\bibnamefont {Gianfreda}}, \bibinfo {author}
  {\bibfnamefont {G.~L.}\ \bibnamefont {Long}}, \bibinfo {author}
  {\bibfnamefont {S.}~\bibnamefont {Fan}}, \bibinfo {author} {\bibfnamefont
  {F.}~\bibnamefont {Nori}}, \bibinfo {author} {\bibfnamefont {C.~M.}\
  \bibnamefont {Bender}}, \ and\ \bibinfo {author} {\bibfnamefont
  {L.}~\bibnamefont {Yang}},\ }\href@noop {} {\bibfield  {journal} {\bibinfo
  {journal} {Nat. Phys.}\ }\textbf {\bibinfo {volume} {10}},\ \bibinfo {pages}
  {394} (\bibinfo {year} {2014})}\BibitemShut {NoStop}%
\bibitem [{\citenamefont {Chang}\ \emph {et~al.}(2014)\citenamefont {Chang},
  \citenamefont {Jiang}, \citenamefont {Hua}, \citenamefont {Yang},
  \citenamefont {Wen}, \citenamefont {Jiang}, \citenamefont {Li}, \citenamefont
  {Wang},\ and\ \citenamefont {Xiao}}]{chang2014parity}%
  \BibitemOpen
  \bibfield  {author} {\bibinfo {author} {\bibfnamefont {L.}~\bibnamefont
  {Chang}}, \bibinfo {author} {\bibfnamefont {X.}~\bibnamefont {Jiang}},
  \bibinfo {author} {\bibfnamefont {S.}~\bibnamefont {Hua}}, \bibinfo {author}
  {\bibfnamefont {C.}~\bibnamefont {Yang}}, \bibinfo {author} {\bibfnamefont
  {J.}~\bibnamefont {Wen}}, \bibinfo {author} {\bibfnamefont {L.}~\bibnamefont
  {Jiang}}, \bibinfo {author} {\bibfnamefont {G.}~\bibnamefont {Li}}, \bibinfo
  {author} {\bibfnamefont {G.}~\bibnamefont {Wang}}, \ and\ \bibinfo {author}
  {\bibfnamefont {M.}~\bibnamefont {Xiao}},\ }\href@noop {} {\bibfield
  {journal} {\bibinfo  {journal} {Nat. Photon.}\ }\textbf {\bibinfo {volume}
  {8}},\ \bibinfo {pages} {524} (\bibinfo {year} {2014})}\BibitemShut {NoStop}%
\bibitem [{\citenamefont {Zhao}\ \emph {et~al.}(2018)\citenamefont {Zhao},
  \citenamefont {Miao}, \citenamefont {Teimourpour}, \citenamefont {Malzard},
  \citenamefont {El-Ganainy}, \citenamefont {Schomerus},\ and\ \citenamefont
  {Feng}}]{zhao2018topological}%
  \BibitemOpen
  \bibfield  {author} {\bibinfo {author} {\bibfnamefont {H.}~\bibnamefont
  {Zhao}}, \bibinfo {author} {\bibfnamefont {P.}~\bibnamefont {Miao}}, \bibinfo
  {author} {\bibfnamefont {M.~H.}\ \bibnamefont {Teimourpour}}, \bibinfo
  {author} {\bibfnamefont {S.}~\bibnamefont {Malzard}}, \bibinfo {author}
  {\bibfnamefont {R.}~\bibnamefont {El-Ganainy}}, \bibinfo {author}
  {\bibfnamefont {H.}~\bibnamefont {Schomerus}}, \ and\ \bibinfo {author}
  {\bibfnamefont {L.}~\bibnamefont {Feng}},\ }\href@noop {} {\bibfield
  {journal} {\bibinfo  {journal} {Nat. Comm.}\ }\textbf {\bibinfo {volume}
  {9}},\ \bibinfo {pages} {981} (\bibinfo {year} {2018})}\BibitemShut {NoStop}%
\bibitem [{\citenamefont {Zhao}\ \emph {et~al.}(2019)\citenamefont {Zhao},
  \citenamefont {Qiao}, \citenamefont {Wu}, \citenamefont {Midya},
  \citenamefont {Longhi},\ and\ \citenamefont {Feng}}]{zhao2019non}%
  \BibitemOpen
  \bibfield  {author} {\bibinfo {author} {\bibfnamefont {H.}~\bibnamefont
  {Zhao}}, \bibinfo {author} {\bibfnamefont {X.}~\bibnamefont {Qiao}}, \bibinfo
  {author} {\bibfnamefont {T.}~\bibnamefont {Wu}}, \bibinfo {author}
  {\bibfnamefont {B.}~\bibnamefont {Midya}}, \bibinfo {author} {\bibfnamefont
  {S.}~\bibnamefont {Longhi}}, \ and\ \bibinfo {author} {\bibfnamefont
  {L.}~\bibnamefont {Feng}},\ }\href@noop {} {\bibfield  {journal} {\bibinfo
  {journal} {Science}\ }\textbf {\bibinfo {volume} {365}},\ \bibinfo {pages}
  {1163} (\bibinfo {year} {2019})}\BibitemShut {NoStop}%
\bibitem [{\citenamefont {Berestetskii}\ \emph {et~al.}(1982)\citenamefont
  {Berestetskii}, \citenamefont {Lifshitz},\ and\ \citenamefont
  {Pitaevskii}}]{berestetskii1982quantum}%
  \BibitemOpen
  \bibfield  {author} {\bibinfo {author} {\bibfnamefont {V.~B.}\ \bibnamefont
  {Berestetskii}}, \bibinfo {author} {\bibfnamefont {E.~M.}\ \bibnamefont
  {Lifshitz}}, \ and\ \bibinfo {author} {\bibfnamefont {L.~P.}\ \bibnamefont
  {Pitaevskii}},\ }\href@noop {} {\emph {\bibinfo {title} {Quantum
  Electrodynamics: Volume 4}}},\ Vol.~\bibinfo {volume} {4}\ (\bibinfo
  {publisher} {Butterworth-Heinemann},\ \bibinfo {year} {1982})\BibitemShut
  {NoStop}%
\bibitem [{\citenamefont {Kawabata}\ \emph {et~al.}(2021)\citenamefont
  {Kawabata}, \citenamefont {Shiozaki},\ and\ \citenamefont
  {Ryu}}]{kawabata2021topological}%
  \BibitemOpen
  \bibfield  {author} {\bibinfo {author} {\bibfnamefont {K.}~\bibnamefont
  {Kawabata}}, \bibinfo {author} {\bibfnamefont {K.}~\bibnamefont {Shiozaki}},
  \ and\ \bibinfo {author} {\bibfnamefont {S.}~\bibnamefont {Ryu}},\
  }\href@noop {} {\bibfield  {journal} {\bibinfo  {journal} {Phys. Rev. Lett.}\
  }\textbf {\bibinfo {volume} {126}},\ \bibinfo {pages} {216405} (\bibinfo
  {year} {2021})}\BibitemShut {NoStop}%
\bibitem [{\citenamefont {Denner}\ \emph {et~al.}(2021)\citenamefont {Denner},
  \citenamefont {Skurativska}, \citenamefont {Schindler}, \citenamefont
  {Fischer}, \citenamefont {Thomale}, \citenamefont {Bzdu{\v{s}}ek},\ and\
  \citenamefont {Neupert}}]{denner2021exceptional}%
  \BibitemOpen
  \bibfield  {author} {\bibinfo {author} {\bibfnamefont {M.~M.}\ \bibnamefont
  {Denner}}, \bibinfo {author} {\bibfnamefont {A.}~\bibnamefont {Skurativska}},
  \bibinfo {author} {\bibfnamefont {F.}~\bibnamefont {Schindler}}, \bibinfo
  {author} {\bibfnamefont {M.~H.}\ \bibnamefont {Fischer}}, \bibinfo {author}
  {\bibfnamefont {R.}~\bibnamefont {Thomale}}, \bibinfo {author} {\bibfnamefont
  {T.}~\bibnamefont {Bzdu{\v{s}}ek}}, \ and\ \bibinfo {author} {\bibfnamefont
  {T.}~\bibnamefont {Neupert}},\ }\href@noop {} {\bibfield  {journal} {\bibinfo
   {journal} {Nat. Comm.}\ }\textbf {\bibinfo {volume} {12}},\ \bibinfo {pages}
  {1} (\bibinfo {year} {2021})}\BibitemShut {NoStop}%
\bibitem [{\citenamefont {Thouless}(1974)}]{thouless1974}%
  \BibitemOpen
  \bibfield  {author} {\bibinfo {author} {\bibfnamefont {D.~J.}\ \bibnamefont
  {Thouless}},\ }\href@noop {} {\bibfield  {journal} {\bibinfo  {journal}
  {Physics Reports}\ }\textbf {\bibinfo {volume} {13}},\ \bibinfo {pages} {93}
  (\bibinfo {year} {1974})}\BibitemShut {NoStop}%
\bibitem [{\citenamefont {Kadanoff}\ and\ \citenamefont
  {Baym}(2018)}]{kadanoff2018quantum}%
  \BibitemOpen
  \bibfield  {author} {\bibinfo {author} {\bibfnamefont {L.~P.}\ \bibnamefont
  {Kadanoff}}\ and\ \bibinfo {author} {\bibfnamefont {G.}~\bibnamefont
  {Baym}},\ }\href@noop {} {\emph {\bibinfo {title} {Quantum statistical
  mechanics: Green’s function methods in equilibrium and nonequilibrium
  problems}}}\ (\bibinfo  {publisher} {CRC Press},\ \bibinfo {year}
  {2018})\BibitemShut {NoStop}%
\end{thebibliography}%

\clearpage

\begin{widetext}
\makeatletter 
\renewcommand{\theequation}{S\arabic{equation}}
\makeatother
\setcounter{equation}{0}

\makeatletter 
\renewcommand{\thesection}{S\arabic{section}} 
\setcounter{section}{0}

\makeatletter 
\renewcommand{\thefigure}{S\@arabic\c@figure}
\makeatother
\setcounter{figure}{0}

\begin{center}
  \textbf{Supplemental Materials for}\\
  \vskip 0.1in
  {\large ``Continuum of Bound States in a Non-Hermitian Model ''}\\
  \vskip 0.1in
  {\small Q.~Wang, C.~Y.~Zhu, X.~Zheng, H.~Xue, B.~Zhang, and Y.~D.~Chong}
\end{center}

\section{Quantization of bound states}
\label{sec:quantization}

Here, we briefly summarize the distinction between discrete and continuous eigenstates, and between bound and free states, in the context of the Hermitian Hamiltonians commonly used in quantum mechanics  \cite{teschl2009mathematical}.

Consider a Hermitian Hamiltonian that is expressible as a differential operator acting on a Hilbert space of wavefunctions.  One example is the Hamiltonian for a non-relativistic particle of mass $m$ in a real potential $V(x)$,
\begin{equation}
  \mathcal{H} = - \frac{1}{2m} \frac{d^2}{dx^2} + V(x),
  \label{Hnonrel}
\end{equation}
which acts on wavefunctions defined in 1D space.

Suppose the problem is defined over a finite spatial interval, $x \in [0,L]$.  The Hamiltonian $\mathcal{H}$ is self-adjoint (i.e., Hermitian), but unbounded.  It can be used to define the resolvent operator $\mathcal{G}(z) = (\mathcal{H} - z)^{-1}$, for $z$ not belonging to the discrete (point) spectrum of $\mathcal{H}$.  Each discrete eigenvalue of $\mathcal{G}(z)$, $g(z)$, has a one-to-one correspondence with an eigenvalue of $\mathcal{H}$, denoted by $E \in \mathbb{R}$, according to $g(z) = (E-z)^{-1}$.  Under these conditions, it can be shown that $G(z)$ is not only bounded but compact; hence, $G(z)$ satisfies the conditions of the spectral theorem (see e.g., Theorem 6.6 in Ref.~\onlinecite{teschl2009mathematical}), implying that its spectrum consists entirely of a countable set of eigenvalues.  We thus conclude that $\mathcal{H}$ has a countable set of discrete eigenvalues (the eigenfunctions are normalizable by assumption).

If $x$ runs over an infinite spatial interval, $G(z)$ need not be compact.  (The time-independent Schr\"odinger equation involving $\mathcal{H}$ is said to be an instance of a ``singular'' Sturm-Liouville problem.)  The spectral theorem then cannot be invoked, so the spectrum of $\mathcal{H}$ is not guaranteed to be entirely discrete.  In general, the spectrum may consist of a continuous part as well as a discrete part.  This is the case for the familiar hydrogen atom potential, which has discrete eigenvalues for $E < 0$ (bound states) and a continuous spectrum for $E > 0$ (free states).

The RAGE theorem (see e.g., Theorem 5.7 in Ref.\cite{teschl2009mathematical}) relates the spectra of self-adjoint Hamiltonians, defined in infinite space, to the behavior of wavefunctions under long time evolution.  Roughly speaking, it states that scattering states (which depart from any local region of space after sufficiently long times) come from the continuous part of the spectrum, whereas bound states (which remain trapped in some local region over long times) come from the discrete part of the spectrum.  This theorem assumes only that $\mathcal{H}$ is self-adjoint (neither $\mathcal{H}$ nor $\mathcal{G}$ need to be compact).

\section{Mapping Continuum Landau Modes to Zeroth Landau Level modes}

The 2D Dirac Hamiltonian, with a symmetric gauge $\vec{A}=[-By,Bx,0]$, is \cite{Peres2006, ZhangGraphene2006, Kentaro2006}
\begin{equation}
\mathcal{H}_D = \begin{pmatrix}
0 & \left(-i\frac{\partial }{\partial x}-By\right)-i\left(-i\frac{\partial }{\partial y}+Bx\right)\\
\left(-i\frac{\partial }{\partial x}-By\right)+i\left(-i\frac{\partial }{\partial y}+Bx\right) & 0
\end{pmatrix}.
\label{DH}
\end{equation}
In the zeroth Landau level (0LL), which has eigenvalue $E=0$, the wavefunction has the form $\begin{pmatrix}\psi_A \\ \psi_B\end{pmatrix}$, where
\begin{align}
  \left[\left(-i\frac{\partial }{\partial x}-By\right)-i\left(-i\frac{\partial }{\partial y}+Bx\right)\right] \psi_B &= 0 \label{DHS1} \\
  \left[\left(-i\frac{\partial }{\partial x}-By\right)+i\left(-i\frac{\partial }{\partial y}+Bx\right)\right] \psi_A &= 0.
  \label{DHS2}
\end{align}
As the two equations are decoupled, we can take either $\psi_A = 0$ or $\psi_B = 0$, depending on the sign of $B$.  In the 2D space is infinite, there is a family of normalizable eigenstates with gaussian spatial envelopes:
\begin{align}
  \psi_A&=0, \qquad \psi_B=Ce^{-B|\mathbf{r}-\mathbf{r}_0|^2/2} \,e^{i\mathbf{q}\cdot\mathbf{r}}, \mathbf{r}_0=[-q_y, q_x]/B, \qquad (B > 0)\\
  \psi_B&=0, \qquad \psi_A=Ce^{+B|\mathbf{r}-\mathbf{r}_0|^2/2} \,e^{i\mathbf{q}\cdot\mathbf{r}}, \mathbf{r}_0=[-q_y, q_x]/B, \qquad (B < 0),
\end{align}
where $C$ is a normalization constant and $\mathbf{q}$ is an arbitrary wavevector.  In the context of graphene, where $\psi_A$ and $\psi_B$ represent wavefunction amplitudes on the two sublattices, the 0LL modes are sublattice polarized \cite{Peres2006, ZhangGraphene2006, Kentaro2006, guinea2010energy, rechtsman2013strain, schomerus2013parity}.

The 0LL modes form an infinite set of gaussian wavepackets, with center position $\mathbf{r}_0$ related to  $\mathbf{q}$ [see Eq.~(3) of the main text].  Since they all have zero energy, they do not form a continuous spectrum (which would violate the principles summarized in Sec.~\ref{sec:quantization}).  At other energies, the other normalizable eigenstates likewise fall into degenerate Landau levels, so $\mathcal{H}_D$ has a purely discrete spectrum of bound states.

In the context of the present work, the operators on the left sides of Eqs.~\eqref{DHS1}--\eqref{DHS2} are equivalent to the non-Hermitian Hamiltonian $\mathcal{H}$ given in Eq.~(1) of the main text, with $s_xs_y = -1$ [for \eqref{DHS1}] or $s_xs_y = 1$ [for \eqref{DHS2}].  Therefore, every 0LL mode is also a zero-energy solution to $\mathcal{H}$.  The non-Hermitian Schr\"odinger equation that we consider is obtained by adding $E \psi_{A/B}$ to the right side of Eq.~\eqref{DHS1} or \eqref{DHS2}.  For any $E \in \mathbb{C}$, we can shift this term to the left side, which is the same as applying a constant shift in the vector potential---a gauge transformation.

For example, for $s_x=-s_y=1$, the time-independent Schr\"odinger equation is
\begin{align}
  \left\{\left[-i \frac{\partial}{\partial x} - By\right]
  - i \left[-i \frac{\partial}{\partial y} + Bx \right] \right\}
  \psi = E \psi  \label{nonhermschrod}
\end{align}
Moving $E$ to the left side gives
\begin{equation}
  \left\{
  \left[-i \frac{\partial}{\partial x} - B\left(y +
    \frac{\mathrm{Re}(E)}{B}\right)\right]
  - i \left[-i \frac{\partial}{\partial y} + B\left(x + \frac{\mathrm{Im}(E)}{B}\right)\right] \right\} \psi = 0,
\end{equation}
which is equivalent to Eq.~\eqref{DHS1} with the displacement
\begin{equation}
  \mathbf{r} \rightarrow \mathbf{r} + \frac{1}{B}\begin{pmatrix}
    \mathrm{Im}(E) \\ \mathrm{Re}(E)
  \end{pmatrix}.
\end{equation}
The eigenstates are the same set of gaussian wavepackets as before, but with the displaced center positions
\begin{equation}
  \mathbf{r}_0= \frac{1}{B}\left[\begin{pmatrix}-q_y \\ q_x
  \end{pmatrix} - \begin{pmatrix}
    \mathrm{Im}(E) \\ \mathrm{Re}(E)
  \end{pmatrix}\right].
\end{equation}

Moreover, we can consider general gauge transformations on the 2D Dirac Hamiltonian.  Suppose we alter Eqs.~\eqref{DHS1}--\eqref{DHS2} by taking
\begin{equation}
  \mathbf{A} = \begin{pmatrix}-By \\ Bx \end{pmatrix}
  \;\;\rightarrow\;\; \mathbf{A}' = \mathbf{A} + \nabla \Lambda(x,y),
\end{equation}
for an arbitrary gauge field $\Lambda(x,y)$.  For any 0LL mode $\psi = \begin{pmatrix}\psi_A \\ \psi_B\end{pmatrix}$ that solves Eqs.~\eqref{DHS1}--\eqref{DHS2}, the gauge transformed Dirac Hamiltonian has an eigenstate $\psi' = \psi e^{-i\Lambda}$ with the same energy $E = 0$.  We can then use the same procedure as above to define a non-Hermitian Hamiltonian.  For example, for the case of $s_x = - s_y = 1$, the gauge transformed non-Hermitian Hamiltonian is
\begin{equation}
  \mathcal{H}' = \left[-i \frac{\partial}{\partial x} - By + \frac{\partial\Lambda}{\partial x} \right]
  - i \left[-i \frac{\partial}{\partial y} + Bx + \frac{\partial\Lambda}{\partial y} \right].
\end{equation}
Its eigenstates can be similarly mapped to a continuous family of 0LL modes, with each $E$ corresponding to an \textit{additional} gauge shift in the 2D Dirac model.  Note that, just as with the Hermitian case, gauge transformations can affect that localization of the eigenstates (e.g., in the symmetric gauge the Landau level modes are localized in 2D, whereas in a Landau gauge they are extended along one direction).

\section{Time-dependent solutions}

In this section, we discuss the time-dependent solutions of the 2D continuum non-Hermitian model.  Let us take $s_x=s_y=1$ (the other cases are straightforwardly related).  The time-dependent Schr\"odinger equation is
\begin{equation}
  i \frac{\partial \psi(x,y,t)}{\partial t}
  =\left[\left(-i\frac{\partial }{\partial x}-By\right)+i\left(-i\frac{\partial }{\partial y}+Bx\right)\right]\psi(x,y,t).
\label{Evolution}
\end{equation}
If we write the initial wavefunction as $\psi(x,y,t=0)=f(x,y)$, the solution is
\begin{equation} 
  \psi(x,y,t) = f(x-t , y-it) \, \exp\left[B(x+iy)t\right],
  \label{solution}
\end{equation}
as can be verified through direct substitution into Eq.~\eqref{Evolution}.  Alternatively, we can write Eq.~\eqref{Evolution} using Wirtinger derivatives:
\begin{equation}
  i \frac{\partial \psi(z,z^*,t)}{\partial t}
  = \left(-2i\frac{\partial}{\partial z^*} + iBz\right) \psi(z,z^*,t).
  \label{EvolutionWirt}
\end{equation}
Then, for an initial wavefunction $\psi(x,y,t=0)=g(z,z^*)$, the solution at other times is
\begin{equation}
  \psi(z,z^*,t) = g(z, z^*-2t) \, e^{Bzt}.
\end{equation}

\begin{figure}
\centering
\includegraphics[width=0.7\textwidth]{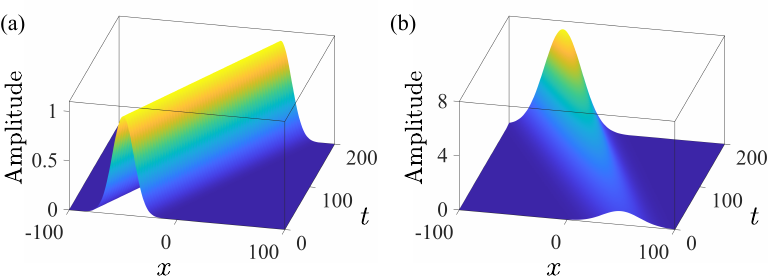}
\caption{Time evolution for two initial gaussian wavepackets, with (a) $\alpha=-0.005$, $\beta=-0.0037$, $x_0=-50$, and $q_y=-0.025$; and (b) $\alpha=-0.0017$, $\beta=-0.0013$, $x_0=50$, $q_y=-0.26$.  The other parameters are $B=-0.005$, $y_0=0$ and $k_x=0$.}
\label{fig:f3}
\end{figure}

In particular, let the initial wavefunction be a gaussian wavepacket:
\begin{align}
  \begin{aligned}
    f(x, y) &= \exp\left[\alpha (x-x_0)^2+iq_x x\right] \exp\left[ \beta (y-y_0)^2+iq_y y\right] \\
  \end{aligned}
\end{align}
Then, using Eq.~\eqref{solution},
\begin{equation}
\begin{aligned} 
  \psi(x,y,t)
  = &\exp\left[\alpha (x-v_0 t-x_0)^2+iq_x x\right]
  \; \exp\left[ \beta (y-y_0)^2+iq_y y\right]\\
  &\times \; \exp\left[(B-\beta-B^2/4\alpha)t^2+(x_0 B+q_y)t\right]\\
  &\times\; \exp\left[i(B-2\beta) yt+(2\alpha x_0+q_y)t+i(2\beta y_0-q_x)t\right],
\end{aligned}
\label{evolution2}
\end{equation}
where $v_0=1-B/2\alpha$. From the first line, we see that the wavepacket retains its gaussian spatial profile for all $t$, with constant widths $\sim \alpha^{-1/2}$ and $\sim \beta^{-1/2}$ in the $x$ and $y$ directions.  The center moves along the $x$ direction with velocity $v_0$, which is positive (negative) if the wavepacket is narrower (broader) than the CLMs.  The overall amplitude can undergo amplification or decay, as indicated by the exponential on the second line.

Fig.~\ref{fig:f3} shows the time evolution of two different initial wavepackets.  The wavepackets are stationary along $y$, so we only show the variation along $x$.  In Fig.~\ref{fig:f3}(a), the parameters are chosen such that the exponent on the second line of Eq.~\eqref{evolution2} vanishes and the wavepacket experiences neither amplification nor damping; moreover, $v_0>0$, so the wavepacket moves in the $+x$ direction.  In Fig.~\ref{fig:f3}(b), the parameters are chosen so that the wavepacket is amplified over time, as well as moving in the $-x$ direction.  In both cases, the width of the wavepacket is constant over time.

\section{Bound State Continua in One Dimensional Models}

In this section, we analyze variants of the non-Hermitian Hamiltonian considered in the main text, and whether they have solutions similar to the continuum Landau modes (CLMs).  We will focus our attention on the 1D case.

First, consider a Hamiltonian that is Hermitian and first-order in momentum:
\begin{equation}
\mathcal{H}=i\frac{\partial }{\partial x} + f(x),
\label{HO-1}
\end{equation}
where $f(x)$ is some real function of $x$. An eigenstate must have the form
\begin{equation}
  \psi(x)=C \exp\left\{-i\int_{-\infty}^x \left[E-f(x')\right]dx' \right\}.
\end{equation}
Since $\mathcal{H}$ is Hermitian, the eigenenergy $E$ must be real.  Hence, the exponent is purely imaginary, so $\psi(x)$ cannot vanish at both $x \rightarrow - \infty$ and $x \rightarrow +\infty$, aside from the trivial solution $\psi(x)=0$.

Next, consider
\begin{equation}
  \mathcal{H} = s_y \frac{\partial}{\partial y} - s_x By.
  \label{Hamiltonain1}
\end{equation}
This can be derived from the continuum 2D model by substituting in the Landau gauge $\mathbf{A} = -By\hat{x}$ and choosing a conserved momentum in the $x$ direction (an additive constant has been omitted).  The 1D lattice model in Eq.~(9) of the main text also reduces to this in the continuum limit.  We can see that there are eigenstates of the form
\begin{equation}
  \psi(y)=C \exp\left[\tau (y-y_0)^2/2 +i q_y y \right],
\end{equation}
where $\tau = - s_xs_y B / 2$.  For $\tau < 0$, these are CLMs.  The corresponding eigenenergies are
\begin{equation}
  E = - \frac{By_0}{s_x} + i \frac{q_y}{s_y}.
\end{equation}

Similar conclusions hold if we multiply the Hamiltonian \eqref{Hamiltonain1} by $i$.  Such a Hamiltonian can be derived from the continuum 2D model by substituting $\mathbf{A} = Bx\hat{y}$ and choosing a conserved momentum along $y$, and also corresponds to the continuum limit of the lattice model described by Eq.~(11) of the main text.

We can obtain other continuous sets of bound states by replacing the $By$ term in Eq.~\eqref{Hamiltonain1} with $By^{n}$, where $n = 3, 5, 7, \dots$  In terms of the 2D model, this is equivalent to specifying a vector potential for a non-uniform out-of-plane magnetic field.  In this case, the solutions have the form $\psi = \exp\left[\tau By^{n+1}+\mathbf{q}\cdot\mathbf{r}\right]$.

\section{Derivation of effective Hamiltonian from lattice model}

In this section, we will go through the derivation of the effective (long-wavelength) Hamiltonian from the lattice model.  For simplicity, we consider the 1D lattice described by Eq.~(10) of the main text (the 2D lattice and the other 1D lattice are handled in a very similar way).  The lattice Hamiltonian has the form
\begin{equation}
\mathcal{H} = \sum_n \Big[ M_n\, a_n^\dagger a_n + t \left(a_{n}^\dagger a_{n-1} + \mathrm{h.c.}\right)\Big],
\label{SH1}
\end{equation}
for $M_n \in \mathbb{C}$ and $t\in\mathbb{R}$.  We take the slowly-varying envelope approximation
\begin{equation}
  |\psi_{k}\rangle = \sum_{n}\exp(i k n) \Psi_{n} a^\dagger_{n}|\varnothing\rangle,
  \label{svea}
\end{equation}
where $\Psi_{n}$ varies slowly relative to the unit cell index $n$, and $|\varnothing\rangle$ is the vacumm state.  Substituting Eq.~\eqref{svea} into Eq.~\eqref{SH1} gives
\begin{equation}
\mathcal{H} |\psi_{k}\rangle = \sum_n \Big[  M_n\, e^{i k n} \Psi_{n} +
 t \left( e^{i k (n-1)} \Psi_{n-1} + e^{i k (n+1)} \Psi_{n+1}   \right)\Big] a^\dagger_{n} \; |\varnothing\rangle.
\label{SH2}
\end{equation}
Now we take
\begin{equation}
  \Psi_{n \pm 1} \approx \Psi_{n} \pm \left.  \frac{\partial \Psi} {\partial x}\right|_n,
\end{equation}
omitting terms of second and higher order (the lattice constant is set to $1$). Eq.~\eqref{SH2} simplifies to
\begin{equation}
  \mathcal{H} |\psi_{k}\rangle
  = \sum_n \Big[ M_n\, \Psi_{n}
    + i 2 t   \sin{k} \frac{\partial \Psi} {\partial x}
    + 2 t \cos{k}\; \Psi_{n} \Big] e^{i k n} a^\dagger_{n}|\varnothing\rangle.
\label{SH3}
\end{equation}
Passing into the continuum limit, we obtain $ H_{k}\Psi(x)=E\Psi(x)$ where
\begin{align}
  H_{k} &= E_{k}^0 + M(x) + i 2t \sin k\; \frac{\partial{}}{\partial{x}},
  \label{SH4} \\
  E_{k}^0 &=2t\cos{k}.
\end{align}

\section{Lattice eigenstates under open and periodic Boundary conditions}

\begin{figure*}[b]
\centering
\includegraphics[width=0.9\textwidth]{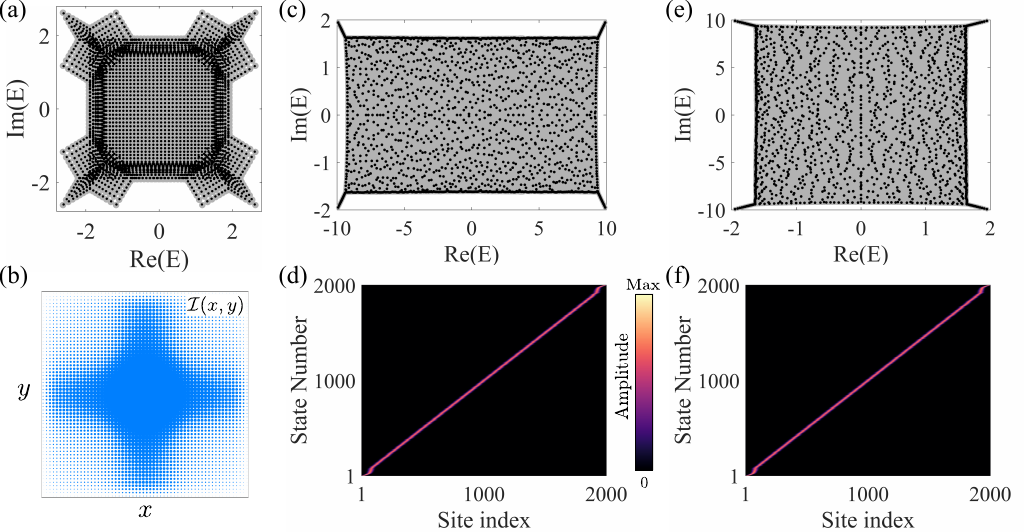}
\caption{(a) Complex energy spectra for the lattice in Fig.~2(a) of the main text, with sample size $60 \times 60$ and $B=0.03$. The black dots indicate the eigenenergies under open boundary conditions (OBC), and the gray dots are eigenenergies plotted for periodic boundary conditions (PBC) as the boundary phases are swept over $[0,2\pi)$. (b) Spatial distribution of the total intensity $\mathcal{I}(x,y)$ for an OBC sample with size $60 \times 60$ and $B=0.03$.  The radius of each blue circle is proportional to $\mathcal{I}(x,y)$.  (c),(e) Complex energy spectra for the 1D lattices depicted in Fig.~3(a) and (d) of the main text with $B=0.01$.  The black dots indicate the eigenenergies under OBC, and the gray dots are the eigenenergies under PBC. (d),(f)  Amplitude distributions for the eigenstates for the two 1D lattices, under OBC.  The lattice size in (c)--(f) is $N=2000$.  }
\label{fig:f5}
\end{figure*}

In this section, we will discuss the eigenvalue spectra for the 2D non-Hermitian lattice presented in the main text, under both open or Dirichlet-like boundary conditions (OBC) and periodic boundary conditions (PBC).  In particular, we demonstrate that the non-Hermitian lattice does not exhibit a non-Hermitian skin effect (NHSE).

Consider the 2D lattice model depicted in Fig.~2(a) of the main text.  For a finite sample of size $60\times60$, the complex eigenvalue spectrum is shown in Fig.~\ref{fig:f5}(a).  Two sets of eigenenergies are plotted here.  The small black dots correspond to the OBC case.  The large gray dots correspond to the PBC case, imposed by connecting the bottom (left) and top (right) boundaries with a phase shift of $k_x$ ($k_y$), and plotting all eigenvalues as $k_x$ and $k_y$ are swept across $[0,2\pi)$.  Evidently, the OBC and PBC eigenenergies occupy the same regions of the complex plane; in models that exhibit a NHSE, the two sets of eigenenergies are expected to be qualitatively different.
  
Fig.~\ref{fig:f5}(b) plots the total intensity $I(x,y)$ for a finite $60\times60$ lattice with OBC.  This quantity is defined as
\begin{equation}
  \mathcal{I}(x,y)=\sum_n^N |\langle x,y|n\rangle|^2,
\end{equation}
where $|n\rangle$ is the $n$-th eigenstate and $(x,y)$ are discrete lattice coordinates.  It can be seen that the total instensity is mostly localized in the bulk, away from the boundaries.  This further implies the absence any NHSE.

The situation is similar for the 1D lattices discussed in the main text.  Fig.~\ref{fig:f5}(c) and (e) show the spectra for the 1D lattices depicted in Fig.~3(a) and (d) of the main text, respectively.  The OBC eigenvalues (black dots) once again occupy the same area of the complex plane as the PBC eigenvalues (gray dots).  (Here, we impose PBC by connecting the first and last sites of the 1D lattice with a hopping $t$, along with a phase shift of $k$ which is swept from $-\pi$ to $\pi$.)  In Fig.~\ref{fig:f5}(d) and (f), we plot the amplitude distributions for the eigenstates of these lattices, under OBC.  No skin modes are found.

\section{Calculation of steady-state solutions}

In the main text, we discuss the effects of applying an external excitation to a non-Hermitian lattice.  In this scenario, we use the time-dependent Schr\"odinger equation with a monochromatic source,
\begin{equation}
  i \frac{\partial |\psi(t)\rangle}{\partial t} = \Big[ H-i\gamma_0 \Big] |\psi(t)\rangle - i\sqrt{2\kappa} e^{-i\omega t} |F\rangle.
\label{EvolutionSd}
\end{equation}
Here, $H$ is the lattice Hamiltonian, $|\psi(t)\rangle = [\psi_1(t), \dots, \psi_N(t)]^T$ is the wavefunction, $N$ is the number of lattice sites, $\gamma_0$ is an additional uniform loss (including outcoupling loss) to prevent blow-up, and $\kappa$ is a uniform coupling coefficient between the lattice and the external excitation.  The excitation frequency is $\omega$, and
\begin{equation}
  |F\rangle = \begin{pmatrix}e^{i\theta_1} \\ \vdots & \\ e^{i\theta_N}
  \end{pmatrix}.
\end{equation}
Each phase $\theta_j$ is drawn independently from the uniform distribution in $[0, 2\pi)$, to represent spatial incoherence.

We seek steady-state solutions $|\psi(t)\rangle = |\psi_0\rangle e^{-i\omega t}$, which satisfy
\begin{equation}
  |\psi_0\rangle = i\sqrt{2\kappa} \Big(H-i\gamma_0-\omega\Big)^{-1} |F\rangle.
\end{equation}
These are calculated using the explicit matrix representation of the lattice Hamiltonian $H$, to produce the results shown in Fig.~3(c) and (f) of the main text.

\section{Behavior of Non-Hermitian Lattices without CLMs}

In this section, we examine the behavior of alternative 1D non-Hermitian lattices that lack CLMs.  When an external excitation is applied, such lattices do not display the rainbow trapping or light funneling behaviors discussed in the main text.

\begin{figure}[b]
\centering
\includegraphics[width=0.8\textwidth]{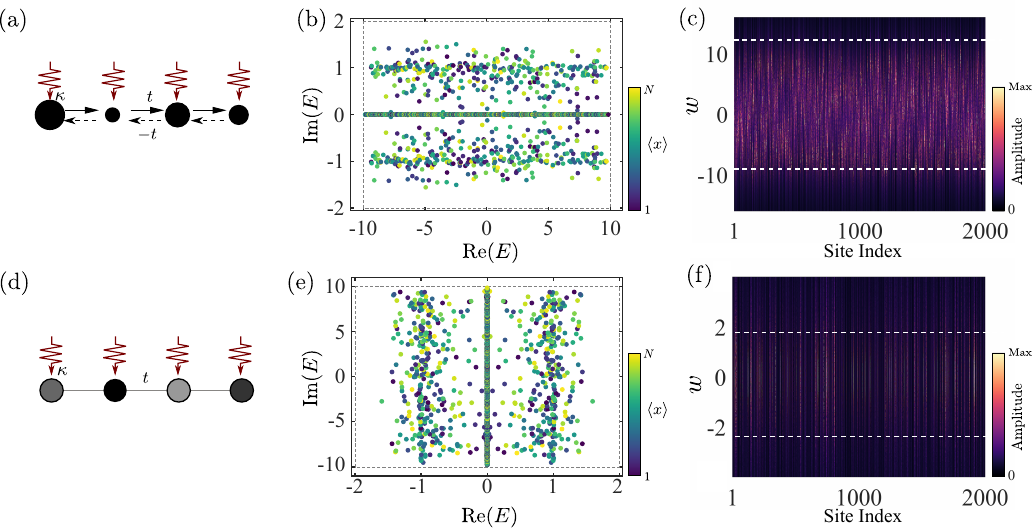}
\caption{Behavior of non-Hermitian 1D lattices with randomly-modulated real or imaginary masses. (a) 1D lattice with nonreciprocal nearest neighbor hoppings $t$ (solid arrows) and $-t$ (dashed arrows).  For each site $j \in [1,N]$, the onsite mass is drawn randomly from $[-BN/2, BN/2]$. (b) Complex energy spectrum, with the color of each dot indicating $\langle x\rangle$.  (c) Wavefunction amplitude induced by an external excitation with uniform amplitude and random phase on each site.  Dashes show the working band $[-BN/2, BN/2]$. (d) 1D lattice with reciprocal nearest neighbor hopping $t$, and onsite masses drawn randomly from $i\times[-BN/2, BN/2]$. (e)--(f) Similar to (b)--(c); the dashes in (f) show the working band $[-2t, 2t]$.  In all subplots, we take $t = 1$, $N = 2000$ and $B=0.01$, along with a uniform damping rate $\Delta m =1.9i$ in (c) and $\Delta m = 9.9i$ in (f).}
\label{fig:f4}
\end{figure}

The first lattice, shown in Fig.~\ref{fig:f4}(a), is similar to the lattice model shown in Fig.~3(a) in the main text, except that in the real onsite mass $m_j$, the spatial gradient is replaced by a random modulation.  We set each $m_j$ to a random value drawn uniformly from $[-BN/2, BN/2]$.  Similar to Fig.~3(a)--(c) in the main text, we take $B=0.01$ and $N=2000$.  Its energy spectrum is plotted in Fig.~\ref{fig:f4}(b).  Unlike Fig.~3(b) in the main text, we see that the eigenenergies are unevenly distributed in the complex plane, with no evident relationship between $E$ and the mean position $\langle x\rangle$.  The response to a spatially incoherent monochromatic excitation is plotted in Fig.~\ref{fig:f4}(c).  Comparing this to Fig.~3(c) of the main text, we see that the excitation excites intensity peaks at numerous random positions throughout the lattice, a characteristic of Anderson localization.  Varying the frequency $\omega$ does not alter the overall spatial distribution of the peaks in a meaningful way.

The second lattice, shown in Fig.~\ref{fig:f4}(c), is similar to Fig.~3(d) except that the imaginary onsite mass is drawn randomly from $i\times[-BN/2, BN/2]$, where $B=0.01$ and $N=2000$.  The spectrum is shown in Fig.~\ref{fig:f4}(e), and the response to a spatially incoherent monochromatic excitation is plotted in Fig.~\ref{fig:f4}(f).  The intensity peaks are distributed randomly across the lattice, and unlike in Fig.~3(f) of the main text, there is no wave funneling behavior.
\clearpage

\end{widetext}

\end{document}